\documentclass[a4paper,11pt,twoside,openright]{article}

\usepackage[utf8x]{inputenc}                               % Ecrire avec les conventions du français.
\usepackage[dvips,pdftex]{graphicx}%\usepackage{graphicx}
\usepackage[left=2.5cm,top=3cm,right=2.5cm,bottom=3cm]{geometry}    % Modification des marges
\usepackage{amsmath,amsthm,amssymb}                             % Utilisation de certains packages de AMS
\usepackage{fancyhdr}
\usepackage{color}

\title{Obliquity of the Galilean satellites:\\
 The influence of a global internal liquid layer}
\author{Rose-Marie Baland$^{a,b}$, Marie Yseboodt$^b$, Tim Van Hoolst$^b$\\
\textit{$^a$Universit\'e de Nantes, UFR des Sciences et des Techniques,}\\ \textit{Laboratoire de Plan\'etologie et G\'eodynamique,}\\ \textit{ 2 chemin de la Houssini\`ere, BP 92208, 44322 Nantes Cedex 3, France.}\\
\textit{$^b$Royal Observatory of Belgium, Ringlaan 3, B-1180 Brussels, Belgium.}\\Email: Rose-Marie.Baland@univ-nantes.fr}
\date{May 2012\\ Paper accepted for publication in Icarus}

\begin{document}

   \maketitle

\section*{Abstract}

The obliquity of the Galilean satellites is small but not yet observed. Studies of cycloidal lineaments and strike-slip fault patterns on Europa suggest that Europa's obliquity is about 1 deg, although theoretical models of the obliquity predict the obliquity to be one order of magnitude smaller for an entirely solid Europa. Here, we investigate the influence of a global liquid layer on the obliquity of the  Galilean satellites. Io most likely has a fully liquid core, while Europa, Ganymede, and Callisto are thought to have an internal global liquid water ocean beneath an external ice shell. We use a model for the obliquity based on a Cassini state model extended to the presence of an internal liquid layer and the internal gravitational and pressure torques induced by the presence of this layer. We find that the obliquity of Io only weakly depends on the different internal structure models considered, because of the weak influence of the liquid core which is therefore almost impossible to detect through observations of the obliquity. The obliquity of Europa is almost constant in time and its mean value is smaller ($0.033-0.044$ deg) with an ocean than without ($0.055$ deg). An accuracy of $0.004$ deg (about $100$ m on the spin pole location at the surface) would allow detecting the internal ocean. The obliquity of Ganymede and Callisto depends more on their interior structure because of the possibility of resonant amplifications for some periodic terms of the solution. Their ocean may be easily detected if, at the measuring time, the actual internal structure model lead to a very different value of the obliquity than in the solid case. A long-term monitoring of their shell obliquity would be more helpful to infer information on the shell thickness.

\section{Introduction}

The Galilean satellites are thought to have internal global liquid layers. In particular, Europa, Ganymede and Callisto likely have an internal global ocean beneath their external ice shell (see Khurana et al. 1998 and Kivelson et al. 2002). It has been shown that observations of longitudinal librations could be used to assess the presence of these liquid layers (Van Hoolst et al. 2008, Baland and Van Hoolst 2010, Rambaux et al. 2011). Another aspect of the rotation state of a synchronous satellite is the time evolution of the orientation of its rotation axis with respect to the normal to its orbit. In general, synchronous satellites are assumed to be locked in a Cassini state, an equilibrium state of a tidally evolved synchronous satellite. This study focuses on the dependence of the obliquity of the Galilean satellites on their internal structure. We assess whether the knowledge of the Galilean satellites obliquity could help confirming the presence of internal global liquid layers in these satellites.

The obliquity of the Galilean satellites is not yet observed but is assumed to be small because it was not detected in gravity measurements (e.g. Anderson et al. 1998). In recent studies, estimates of the obliquity of Europa have been obtained from observations of the cycloidal  lineaments and strike-slip faults (e.g. Rhoden et al. 2010,  2011). They suggest an order of magnitude from a few tenths of a degree to a degree for Europa's obliquity. Observations of the obliquity and of the librations for Europa and maybe for Ganymede are expected in the near future, thanks to radar echoes from the satellite's surfaces observed from two separate Earth stations (Margot 2011). This technique has been proved to be efficient, since it allowed the detection of a forced libration and of a non-zero obliquity for Mercury, which confirmed that this planet has a molten core and is in a Cassini state (Margot et al. 2007).

In this paper, we apply the Cassini state model for a synchronous satellite with a liquid ocean, developed and applied to Titan in Baland et al. (2011), to the Galilean satellites. The paper is organized as follows. In the second section, we briefly introduce the Cassini state model for an entirely solid satellite, based on the model of Bills (2005), and we compare his results with ours. In the third section, we extend the model taking into account the presence of an internal liquid layer and the resulting gravitational and pressure torques between the different layers of a satellite. In the same section, we present the results for the liquid layer case for the four satellites and we assess the possibility to detect a fluid layer and to constrain the satellite's interior structure from obliquity observations. The last section contains a discussion of our results and the conclusions.

\section{The Cassini state for an entirely solid satellite}
\subsection{Theory}

Let us first consider that the Galilean satellites are entirely solid. If principal axis rotation is assumed, that is to say that a possible wobble is neglected, the angular momentum equation governing the obliquity of a synchronous satellite is (see Bills (2005) and the demonstration given in appendix \ref{Append})
\begin{eqnarray}
  n C \frac{d\hat s}{dt}&=&n \kappa (\hat s \wedge \hat n),\nonumber \\
\label{rigidcase}\kappa&=&\frac{3}{2}MR^2(-C_{20}+2C_{22})n=\frac{3}{2}(C-A)n,
\end{eqnarray}
where $\hat s=(s_x,s_y,s_z)$ and $\hat n=(n_x,n_y,n_z)$ are the unit vectors along the rotation axis and the normal to the orbit, expressed in cartesian coordinates $(x,y)$ of the Laplace plane and $z$ along the normal to it. The Laplace plane is the mean orbital plane of the satellite with respect to which the orbital precession is defined and is here considered as the equatorial plane of an inertial reference frame. The mean motion, equal to the mean rotation rate, is denoted by $n$ and the mass and the radius of the satellite by $M$ and $R$. $C_{20}$ and $C_{22}$ are the second-degree gravity field coefficients of the satellite, which can be expressed in terms of $A<B<C$, the principal moments of inertia. The left-hand member of Eq. (\ref{rigidcase}) is the variation of the angular momentum while the right-hand member is the gravitational torque (averaged over the orbital period) exerted by the planet Jupiter that tends to align the rotation axis with the orbit normal. A perfect permanent alignment is impossible to achieve because of the continuous precession, leading to a non-zero obliquity. Following Bills (2005), this vectorial equation (\ref{rigidcase}) is projected on the Laplace plane, 
\begin{eqnarray}\label{rigidprojected}
 \frac{d S}{dt}=I \frac{\kappa}{C} (N-S)
\end{eqnarray}
with $S=s_x+I s_y$ and $N=n_x+I n_y$, $I=\sqrt{-1}$. This equation is correct up to the first order in small obliquity, eccentricity and orbital inclination.

Equation (\ref{rigidprojected}) allows the study of the classical Cassini state, where the precession rate $\dot \Omega$ is constant, as well as the study of the generalization to a multi-frequency node precession  (Bills 2005). In the latter case, the orbital precession can be written as a series expansion
\begin{eqnarray}\label{precession}
N=\sum_j\sin{i_j}\, e^{I (\dot\Omega_j t+ \gamma_j-\pi/2)},
\end{eqnarray}
where $i_j$ are the inclination amplitudes associated with the orbital node precession frequencies $\dot\Omega_j$ and the phases $\gamma_j$. The actual orbital inclination $i$, which is the angle between the normal to the Laplace plane and the normal to the orbit, is not constant over time and is given, at the first order, by 
\begin{eqnarray}\label{incl}
i&\simeq&\|N\|.
\end{eqnarray}
The parameter $\dot\Omega_1$ is the main precession rate of the orbital plane, with respect to the Laplace plane, which is the only frequency considered in the classical Cassini state. In that case, the orbital inclination $i$ stays constant in time and is equal to the inclination amplitude $i_1$.

The forced precession of the rotation axis induced by the precession of the orbital plane (\ref{precession}) is given by the forced solution of equation (\ref{rigidprojected}):
\begin{eqnarray}\label{forcedsol}
S&=&\sum_j \sin{(i_j+\varepsilon_j)}\, e^{I (\dot\Omega_j t+ \gamma_j-\pi/2)},
\end{eqnarray}
where, correct up to the first order in $\varepsilon_j$ and $i_j$, the obliquity amplitudes $\varepsilon_j$ associated with the frequencies $\dot\Omega_j$ are given by
\begin{eqnarray}\label{rigidfirstorder}
\varepsilon_j=-\frac{i_j\dot\Omega_j}{(\omega_f+\dot\Omega_j)},
\end{eqnarray}
in which
\begin{eqnarray}
\label{wf} \omega_f=\frac{\kappa}{C}
\end{eqnarray}
is the frequency of the free spin precession obtained by averaging the torque over a time scale larger than the precession time scale, or equivalently by setting $N$ to zero. %The period corresponding to this frequency is noted $T_f$. 
In the series expansion of Eq. (\ref{precession}), the inclination amplitudes $i_j$ are ordered such that their magnitude decreases with increasing subscript $j$. The obliquity amplitudes $\varepsilon_j$ do not necessarily follow this trend, because $i_j$ is multiplied by $\dot\Omega_j$ in the numerator of the solution (\ref{rigidfirstorder}) and because the denominator $(\omega_f+\dot\Omega_j)$ may be small and lead to large value of $\varepsilon_j$. If $\varepsilon_j$ is significantly amplified due to $\omega_f$ being close to $(-\dot\Omega_j)$, one can speak of a resonance between the free mode and the forcing frequency. In the classical Cassini state, the obliquity is given by $\varepsilon=\varepsilon_1$ and is constant, but in the generalized Cassini state, the obliquity is time-variable and is given by (see also Bills and Nimmo 2011a)
\begin{eqnarray}\label{obl}
\varepsilon&\simeq&\|S-N\|
\end{eqnarray}
In the classical Cassini state, the rotation axis, the normal to the orbit and the normal to the Laplace plane remain in the same plane. In the generalized Cassini state, the rotation axis presents a time-variable deviation with respect to the plane defined by the two other axes. An expression of this deviation can easily be obtained from Eqs. (\ref{precession}) and (\ref{forcedsol}) (see Eq. (9) of Baland et al. 2011). The deviation is close to zero when none of the obliquity amplitudes are amplified by a resonance, or equivalently when the rotation state is close to the classical Cassini state, while the obliquity is non-zero and close to $\varepsilon_1$. Deviation and obliquity variations depend both on obliquity amplitudes $\varepsilon_j$. The trajectories of $N$ and $S$ on the Laplace place are seen as the composition of circular motions (see Eqs. (\ref{precession}) and (\ref{forcedsol}), see also Fig. 3 of Baland et al. 2011). It then follows from the geometry of the problem that the time periodic variations of the deviation around zero have amplitudes of the same order of magnitude as the amplitude of the time periodic variation of the obliquity around $\varepsilon_1$, but that they are out of phase, since the deviation is maximum when the obliquity is close to its mean value. Therefore, study both the variations of the obliquity and of the deviation is somewhat redundant. In the following, we will focus on the numerical results for the obliquity and their interpretation in terms of internal structure. Nevertheless, future observations of the rotation state of the Galilean satellites should be interpreted in term of internal structure by taking into account the obliquity and the deviation in the case of resonant amplification, as it was done for Titan in Baland et al. (2011). The measured deviation could help to check the consistency of the model and, if more than one resonant amplification is able to explain a large measured obliquity, to chose among them.

\subsection{Numerical results}

We calculate the obliquity of the Cassini state for the Galilean satellites assuming that they are entirely solid. The numerical values of the radius $R$ and of the mean density $\bar \rho$, which give the mass $M$, and of the gravity field coefficients $C_{20}$ and $C_{22}$, according to Schubert et al. (2004), are given in Table 1 for the four Galilean satellites. The normalized mean moment of inertia $I$ can be obtained using the extension of Radau's equation to synchronous triaxial bodies (see Van Hoolst et al. 2008). Table 1 also gives the mean motion $n$ according to Lainey et al. (2006) and an estimation of the free precession period $T_f=2\pi/\omega_f$ using Eq. (\ref{wf}) and the approximation $C\simeq I$. For Io, Europa and Callisto, $T_f$ is in a satisfying agreement with previous studies based on the Hamiltonian mechanics (Henrard 2005b gives $0.436$ years for Io, Henrard 2005a gives $3.603$ years for Europa and Noyelles 2009 gives $203.600$ years for Callisto). The parameters of the orbital precession series expansion (orbital inclination amplitudes $i_j$, precession frequencies $\dot\Omega_j$ and their corresponding period $T_j$, and phase angles $\gamma_j$) have been taken from Lainey et al. (2006) and are given respectively in Tables 2, 3, 4 and 5 for Io, Europa, Ganymede and Callisto. The last column of Tables 2-5 gives the obliquity amplitudes $\varepsilon_j$ computed with Eq. (\ref{rigidfirstorder}). 

Significant resonant amplifications occur in the solid case for the subscript $j=5$ for Io, $j=3$ for Ganymede, and $j=2$ for Callisto, but there is no significant resonance for a solid Europa. As a result, the solid case obliquity of Europa remains close (at the 6 percent level) to the constant value of the classical Cassini state ($\varepsilon_{1}=0.055^\circ$). This is one order of magnitude lower than the estimate of $0.32^\circ-1.35^\circ$ obtained from the observations of cycloidal lineaments (Rhoden et al. 2010) and the one of $1^\circ-1.25^\circ$ from strike-slip fault patterns (Rhoden et al. 2011). The obliquity of Io oscillates essentially between $(\varepsilon_{1}-\varepsilon_{5})=0.0015^\circ$ and $(\varepsilon_{1}+\varepsilon_{5})=0.0027^\circ$. Obliquities of Ganymede and of Callisto evolve respectively in the ranges $[0^\circ,0.067^\circ]$ and $[0.046^\circ,0.244^\circ]$.

The time evolution of the orbital inclination $i$ and of the obliquity $\varepsilon$, computed with Eqs. (\ref{incl}) and (\ref{obl}), is illustrated in Fig. 1 for the four Galilean satellites. We have chosen J2000 as the time origin. The width of the time spans are the same as in Bills (2005) to facilitate the comparison with his figure 3 which represents also the time evolution of $i$ and $\varepsilon$ of the four satellites computed from the same equations but with different ephemerides. A visual comparison indicates clearly that we are far from agreement, both for the orbital inclination and the obliquity.

An obvious cause for this difference is the error which has lead to the absence of the coefficient $C_{22}$ in the angular momentum equation of Bills (2005) (see his equation 46 which is the equivalent of equation \ref{rigidfirstorder} of the present paper). This changes the numerical values of the frequency $\omega_f$ and so of the obliquity amplitudes $\varepsilon_j$ by affecting possible resonant amplifications. This effect alone, however, is not sufficient to explain the differences between our curves for the obliquity and those of Bills. Moreover, it does not affect the inclinations. 

Part of the difference may also be due to a possible different use of ephemerides. Bills (2005) used the ephemerides of Lieske (1998) to calculate the parameters of the orbital precession series expansion of Eq. (\ref{precession}). As those ephemerides are expressed with respect to the equatorial plane of Jupiter, this plane was most likely considered as the inertial plane of reference, instead of the Laplace plane. Here, we use the synthetic representation of ephemerides of Lainey et al. (2006), also expressed with respect to the equatorial plane of Jupiter, but which could easily be corrected for the difference with respect to the Laplace plane. In the expansions for the orbital precessions given in Lainey et al. (2006), we simply neglect a term with precession frequency equal to zero, which gives the orientation of the Laplace plane with respect to the equatorial plane of Jupiter. Introducing that term in expansion (\ref{precession}), we find orbital inclinations that are in a better agreement with the ones of Bills (2005). However, the obliquity is not affected by the choice of the reference plane. 

The remaining differences might be caused by possible differences in the phases $\gamma_j$ used in solution (\ref{forcedsol}) and, more generally, by the differences between the different ephemerides used that can affect the possible resonant amplifications, particularly in the case of Io for which the orbital inclination behavior presented in Bills (2005) seems to be the combination of a few terms in the expansion (\ref{precession}).

\section{The Cassini state model for a satellite with a global liquid layer}
\subsection{Theory}

We develop a generalized Cassini state model for a generic satellite made up of three uniform and homogeneous layers: a rigid ice shell ($sh$), a liquid water ocean ($o$) and a rigid interior ($in$) as illustrated in Fig. 2. This definition is adequate for Europa, Ganymede and Callisto. For Io, the shell will be replaced by a rigid rock mantle ($m$), the water ocean by a fluid core ($f$), and no solid interior will be considered.

For the generic satellite, the orientation of each layer will in principle be described by one angular momentum equation. An external gravitational torque exerted by the central planet applies on each layer, as well as torques resulting from gravitational and pressure interactions between the layers. Under the assumption that the liquid is in hydrostatic equilibrium, the total torque on the liquid layer vanishes (Baland et al. 2011). Therefore we only consider the orientation of the unit vectors along the rotation axes of the shell ($\hat s_{sh}$) and the interior ($\hat s_{in}$), which can have different obliquities ($\varepsilon_{sh}$ and $\varepsilon_{in}$) with respect to the normal to the orbit $\hat n$ (see Fig. 2). The system then reduces to two angular momentum equations, which can be written correct up to the first order in obliquity and inclination as 
\begin{eqnarray}
\label{reduce1} C_{sh} \frac{dS_{sh}}{dt}&=&I \kappa'_{sh}(N-S_{sh})-I K (S_{in}-S_{sh}),\\
\label{reduce2} C_{in} \frac{dS_{in}}{dt}&=&I \kappa'_{in}(N-S_{in})+I K(S_{in}-S_{sh}),
\end{eqnarray}
where $S_{sh}$ and $S_{in}$ are the projections on the Laplace plane of $\hat s_{sh}$ and $\hat s_{in}$, respectively. $C_{sh}$ and $C_{in}$ are the polar moment of inertia of the two solid layers. $K$ represents the strengths of the internal gravitational torque between the two solid layers, corrected for the internal pressure torque. $\kappa'_{sh}$ and $\kappa'_{in}$ are the strength of the external torques exerted by the planet on the shell and on the interior, also corrected for the pressure effect. We have (Baland et al. 2011)
\begin{eqnarray}
 \kappa'_{sh}&=&\frac{3}{2}n[(C_{sh}-A_{sh})+(C_{o,t}-A_{o,t})],\\
 \kappa'_{in}&=&\frac{3}{2}n[(C_{in}-A_{in})+(C_{o,b}-A_{o,b})],\\
 K&=&-[(8 \pi\, G)/(5 n)] [(C_{in}-A_{in})+(C_{o,b}-A_{o,b})]\times\nonumber\\
 && [\rho_{sh}(\alpha_{sh}-\alpha_o+\beta_{sh}/2-\beta_o/2)+\rho_o (\alpha_o+\beta_o/2)],
\end{eqnarray}
where $A_{sh}$ and $A_{in}$ are the smallest principal moments of inertia of the two layers. $\alpha$ and $\beta$ are the polar and equatorial flattenings of the differents layers, defined as the relative differences $((a+b)/2-c)/((a+b)/2)$ and $(a-b)/a$, respectively, with $a>b>c$ the radii in the direction of the principal axes of the layer considered. The subscripts $(o,t)$ and $(o,b)$ refer to the mass of the ocean aligned with the shell and the interior, respectively, or equivalently, to a top ocean and a bottom ocean above and beneath an arbitrarily chosen sphere inside the ocean (see Fig. 2). The terms in the torques strengths that depend on the density of the ocean reflect the pressure effect of the ocean. The moments of inertia $A_l$ and $C_l$, and the difference $(C_l-A_l)$ of any layer $(l)$, are given by
\begin{eqnarray}
\label{Al}A_l&=&\frac{8\pi}{15}\rho_l\left[R_l^5\left(1-\frac{1}{3}\alpha_l-\frac{1}{2}\beta_l\right)-R^5_{l-1}\left(1-\frac{1}{3}\alpha_{l-1}-\frac{1}{2}\beta_{l-1}\right)\right]\\
\label{Cl}C_l&=&\frac{8\pi}{15}\rho_l\left[R_l^5\left(1+\frac{2}{3}\alpha_l\right)-R^5_{l-1}\left(1+\frac{2}{3}\alpha_{l-1}\right)\right]\\
\label{CmA} (C_l-A_l)&=&\frac{8\pi}{15}\rho_l\left[R_l^5\left(\alpha_l+\frac{1}{2}\beta_l\right)-R^5_{l-1}\left(\alpha_{l-1}+\frac{1}{2}\beta_{l-1}\right)\right],
\end{eqnarray}
with the subscript $(l-1)$ referring to the layer located beneath layer $l$ (e.g. Van Hoolst et al. 2008).

For the orbital precession (\ref{precession}), the system of Eqs. (\ref{reduce1}-\ref{reduce2}) has a forced solution for the obliquity amplitudes of both the shell and the interior:
\begin{eqnarray}
\label{Ssh}S_{sh}&=&\sum_j (i_j+\varepsilon_{j,sh}) e^{I (\dot \Omega_j t+ \gamma_j-\pi/2)},\\
\label{Sin}S_{in}&=&\sum_j (i_j+\varepsilon_{j,in}) e^{I (\dot \Omega_j t+ \gamma_j-\pi/2)},\\
\label{solution}
\varepsilon_{j,sh}&=&\frac{i_j \dot\Omega_j (K(C_{sh}+C_{in})-C_{sh} \kappa'_{in} -C_{in}\, C_{sh} \,\dot\Omega_j)}{C_{in}C_{sh}(\omega_{+}+\dot\Omega_j)(\omega_{-}+\dot\Omega_j)},\\
\label{solution2}\varepsilon_{j,in}&=&\frac{i_j \dot\Omega_j (K(C_{sh}+C_{in})-C_{in} \kappa'_{sh}-C_{in}\, C_{sh} \,\dot\Omega_j)}{C_{in}C_{sh}(\omega_{+}+\dot\Omega_j)(\omega_{-}+\dot\Omega_j)}.
\end{eqnarray}
with the two free mode frequencies given by
\begin{eqnarray}
\label{omegapm}\omega_{\pm}&=&-(Z \pm \sqrt{\Delta})/(2 C_{in} C_{sh}),\\
Z&=&K (C_{in}+C_{sh})-C_{sh} \kappa'_{in} -C_{in} \kappa'_{sh}, \nonumber\\
\label{freemodes}\Delta&=&-4 C_{in} C_{sh} (-K (\kappa'_{in}+\kappa'_{sh})+\kappa'_{in} \kappa'_{sh} )+Z^2. \nonumber
\end{eqnarray}
The frequency $\omega_+$ corresponds to a mode where the shell and the interior precess in phase with each other, while they are out of phase for $\omega_-$. 

As in the solid case, a significant amplification of the obliquity amplitudes $\varepsilon_{j,sh}$ and $\varepsilon_{j,in}$ might occur if $\omega_{+}$ or $\omega_{-}$ is close enough to ($-\dot \Omega_j$). Nevertheless, very large obliquity amplitudes of the order of several tens of degrees do not make sense, since the analytical solution (17-18) was derived from Eqs. (9-10) which are only valid for small angles. This, however, is not problematic since the obliquity of the Galilean satellites is known to be less than about $10$ deg (as can be estimated from the observed gravitational coefficients $C_{21}$ and $S_{21}$), in which case the small angles assumption leads at most to an error of a few percent for obliquity amplitudes of a few degrees. Any internal structure model that would reach very high obliquity values can be excluded and we can consider that our solution is valid for application to the Galilean satellites.. 

\subsection{Io}

We consider that Io consists of two homogeneous layers: a solid silicate mantle over a liquid iron-sulfur core. The core density is chosen in the range $[5150,8000]$ kg/m$^3$, depending on the content in light elements (see Anderson et al. 2001b, Spohn et al. 1997 and Schubert et al. 2004). The interior structure is characterized by four parameters which are the radius and density of the core and of the mantle ($R_f$ and $R_m$, $\rho_f$ and $\rho_m$), and is constrained by the mean radius $R$, the mass $M$, and the mean moment inertia $I$ given in Tab. 1. Obviously, the radius of the mantle is equal to the radius of the satellite, so that $R_m=R$. With the two remaining constraints ($M$ and $I$), the solution for the three remaining parameters ($\rho_f$, $\rho_m$ and $R_f$) is not unique. However, given the possible range of core densities $\rho_f$, we can find limit values for the density and the thickness of the mantle. Since the mantle thickness is given by $h_m=R-R_f$, a core density $\rho_f$ of $5150$ kg/m$^3$ corresponds to a mantle thickness $h_m$ of $877.6$ km and a mantle density $\rho_m$ of $3265.1$ kg/m$^3$. If $\rho_f$ is $8000$ kg/m$^3$, then $h_m=1164.1$ km and $\rho_m=3306.8$ kg/m$^3$.

We want to assess the effect of the mantle thickness on the obliquity. So, we consider values for $h_m$ which are equally spaced by 0.5 km, between the two limit values given before ($877.6$ and $1164.1$ km) and we compute $\rho_m$ and $\rho_f$ from the mass and moment of inertia constraints. Next, we solve Clairaut's equation to get polar and equatorial flattenings of the core for the different interior models (see e.g. Van Hoolst et al. 2008) and we compute the moment of inertia difference $(C-A)$ of the different layers with Eq. (\ref{CmA}). The solution for the obliquity amplitudes of the mantle can then be determined from Eq. (\ref{solution}) by taking the limit of no solid interior. We then have
\begin{eqnarray}
\label{solutionIo}
\varepsilon_{j,m}&=&-\frac{i_j\dot\Omega_j}{(\omega_{+}+\dot\Omega_j)},\\
\omega_{+}&=&\frac{\kappa'_{m}}{C_m}\\
\label{kappamp}\kappa'_{m}&=&\frac{3}{2}n[(C_{m}-A_{m})+(C_{f,t}-A_{f,t})]=\frac{3}{2}n(C-A)=\kappa.
\end{eqnarray}
The strength of the external torque exerted on the mantle, $\kappa'_m$, is equal to the strength $\kappa$ of the external torque on the whole satellite (see Eq. \ref{rigidcase} and \ref{kappamp}) because of the lack of an internal solid layer and because of the pressure correction on the external torque. As a result, the rotation axis of Io's mantle precesses as a rigid body with a polar moment of inertia equal to $C_m$ but with a moment of inertia difference $(C-A)$ equal to the one of the entire body. Solution (\ref{solutionIo}) can also be written as
\begin{eqnarray}
\label{solutionIo2}
\varepsilon_{j,m}&=\varepsilon_{j}&\frac{\omega_f+\dot\Omega_j}{C\omega_f/C_m+\dot\Omega_j}.
\end{eqnarray}
with $\varepsilon_{j}$ the solution for the solid case (\ref{rigidfirstorder}) and 
\begin{eqnarray}
% C_m&=&\frac{8\pi}{15}\rho_{m}\left[R^5\left(1+\frac{2}{3}\alpha_{m}\right)-R_{f}^5\left(1+\frac{2}{3}\alpha_{f}\right)\right]\\
\label{Cmapprox}C_m&\approx& \frac{8\pi}{15}\rho_{m}[R^5-(R-h_{m})^5]
\end{eqnarray}
because of Eq. (\ref{Cl}) with $\alpha_m$ and $\alpha_f\ll 1$.

Equation (\ref{Cmapprox}) shows that $C_m$ increases with increasing mantle thickness $h_m$. $C_m$ ranges from $94$ to $98.5\%$ of the value of $C$. As a consequence, the period $T_{+}$ corresponding to the free mode frequency $\omega_+$, in the range of $0.388-0.406$ years, is slightly smaller than the period $T_f=0.41$ years of the free precession of the solid case  and increases with increasing $h_m$ (Fig. 3). The possible range of obliquity amplitudes of the mantle $\varepsilon_{j,m}$ are, for each subscript $j$, very close to the obliquity amplitudes of the solid case $\varepsilon_{j}$ (Fig. 4) and slightly increase with increasing $h_m$ (see Fig. 5 for $\varepsilon_{1,m}$ and $\varepsilon_{5,m}$). In particular, for $j=1,2,3,4$ or $6$, $\dot\Omega_j$ can be neglected compared to $C\omega_f/C_m$ and the corresponding obliquity amplitudes are from $1.5\%$ to $6\%$ smaller in magnitude than in the solid case. The case $j=5$ is somewhat different because $T_+$ is rather close to $-T_5(=0.68$ years$)$, the opposite of the period of the fifth orbital precession frequency $\dot \Omega_5$ defined in Eq. (\ref{precession}) and given in Tab. 2. Therefore, as in the entirely solid case, only $\varepsilon_{5,m}$ competes with $\varepsilon_{1,m}$, $\dot\Omega_5$ cannot be neglected in front of $C\omega_f/C_m$ and $\varepsilon_{5,m}$ is from $3\%$ to $13\%$ smaller than for the solid case (see also Tab. 6).

Since the obliquity amplitudes of the fluid core case are close to those of the entirely solid case, the actual time-dependent obliquity of both cases are similar, as can be seen in Fig. 6 where we show the obliquity of the solid case and of the internal structure models corresponding to minimal and maximal values of $\varepsilon_{1,m}$. Time variations are mainly due to the amplified $\varepsilon_{5,m}$ and $\varepsilon_{5}$. A very accurate measurement of the obliquity would be needed to detect a fluid core, especially if the mantle is thick. The error on the measurement would then have to be of the order of $5\,10^{-6}$ degrees (corresponding to an error of the order of $0.16$ m on the location of the rotation axis at the surface). Constraining the mantle thickness to a ten-kilometer level will be even more challenging and would require a precision improved by a factor ten. We do not think that such an accuracy could be achieved in the near future, since, for instance, the accuracy on future Europa's obliquity measurement from the Earth is expected to be of the order of $0.1^\circ$ (Margot 2011) and since no space mission dedicated to Io is planned. 

\subsection{Europa}

Anderson et al. (1998) concluded from Galileo gravity measurements that Europa is differentiated into a metallic core, a rocky mantle and a liquid/ice water layer. the detection of an induced magnetic field (Khurana et al. 1998) indicates that a liquid layer would exist between the ice surface and the rocky interior. Therefore, we consider that Europa is made of four homogeneous layers, from the center to the surface: an iron-sulfur core (c), a rock mantle (r), a liquid water ocean (o), and a water ice shell (sh). The mantle and the core together form a solid layer to which we refer as the interior (in). Since the mean radius $R$, given in Tab. 1, is also the radius of the ice shell $R_{sh}$, we have only two data (the mass $M$ and the mean moment of inertia $I$) to constrain seven parameters which are the radii $R_c$, $R_r$ and $R_o$ and the densities $\rho_c$, $\rho_r$, $\rho_o$ and $\rho_{sh}$, and the solution for the internal structure is not unique. As in Baland and Van Hoolst (2010), we consider a representative range of possible equally spaced values for five parameters: $h_{sh}$ can be equal to $5, 7.5,..., 15$ km or $20, 30,..., 100$ km; $R_r$ can be $1400, 1425,...,1475$ km; $R_c$ is $100, 200,..., 800$ km; while $\rho_{sh}$ and $\rho_{o}$ can be equal to $800,900,..., 1200$ kg/m$^3$. We then compute $\rho_r$ and $\rho_c$ from $M$ and $I$ for any chosen set of values. We only retain solutions for which $3000\leq\rho_r\leq3800$ kg/m$^3$, $5150\leq\rho_c\leq 8000$ kg/m$^3$, $R_o>R_r$, and $\rho_{sh}\leq\rho_{o}$. This procedure provides $1434$ solutions for the internal structure.

In order to easily identify the possible resonant amplification that might occur with some negative periods $T_j$ of the orbital precession, we have plotted $T_+$ as a function of $T_-$ to (see fig. 4). The negative period $T_4=-7.43$ years is the closest to the range of values of $T_+$ (from $3.1$ to $3.7$ years) while $T_5=-0.68$ years is the closest to the range of $T_-$ (from $0.01$ to $0.34$ years). Nevertheless, the resonant amplifications for $j=4$ or $j=5$ are not important enough to allow the corresponding obliquity amplitudes of the shell and of the interior to compete with those corresponding to the main frequency precession $\varepsilon_{1,sh}$ and $\varepsilon_{1,in}$ (Fig. 4). As a result, the solution (\ref{Ssh}-\ref{Sin}) is dominated by the first term of its series expansion, as for an entirely solid Europa, and the orientation of the rotation axis of the different layers is close to the coplanarity with the normal to the orbit and the normal to the Laplace plane. 

For any given frequency $\dot \Omega_j$, the obliquity amplitudes of the interior $\varepsilon_{j,in}$ differ from the obliquity amplitudes of the shell $\varepsilon_{j,in}$ (Fig. 4), but are close to the obliquity amplitudes of the solid case, $\varepsilon_{j}$. These facts are due to the different torques exerted by Jupiter on the shell and on the interior and to the internal coupling being too weak to restore a perfect alignment and can be understood quite intuitively. First, consider the interior perfectly decoupled from the shell. %($K=0$, $\rho_s=\rho_o=0$) 
It would have obliquity amplitudes given by (see Eqs. \ref{rigidfirstorder} and \ref{wf})
\begin{eqnarray}
\label{solutionInt}
\varepsilon^{dec}_{j,in}&=&-\frac{i_j\dot\Omega_j}{(\omega_{in}+\dot\Omega_j)},\\
\omega_{in}&=&\frac{3}{2}n\frac{(C_{in}-A_{in})}{C_{in}},
\end{eqnarray}
and would react to the external torque similarly as the entire satellite in the solid case since it contains most of the satellite's mass (e.g., for a typical internal structure model, we have $C_{in}\simeq 0.85 C$, $(C_{in}-A_{in})\simeq 0.84 (C-A)$, and $\omega_{in}\simeq0.98 \omega_f$, $\omega_f$ being the free mode frequency of the solid case, see Eq. \ref{wf}). The shell %($K=0$, $\rho_o=\rho_r=\rho_c=0$)
would have obliquity amplitudes
\begin{eqnarray}
\label{solutionsh}
\varepsilon^{dec}_{j,sh}&=&-\frac{i_j\dot\Omega_j}{(\omega_{sh}+\dot\Omega_j)},\\
\omega_{sh}&=&\frac{3}{2}n\frac{(C_{sh}-A_{sh})}{C_{sh}}
\end{eqnarray}
and would differ more from the solid $\varepsilon_{j}$ than $\varepsilon^{dec}_{j,in}$ since the difference of $\omega_{sh}$ with $\omega_f$ is larger than the difference of $\omega_{in}$ with $\omega_{f}$ ($\omega_{sh}\simeq1.07 \omega_f$ for the same particular model). If we now add the internal coupling, the interior is less likely to be disturbed from the decoupled case and $\varepsilon_{j,in}$ tends to stay closer to $\varepsilon_{j}$ than $\varepsilon_{j,sh}$, because the inertia of the interior $C_{in}$ is about 6.5 times larger than the one of the shell $C_{sh}$. A more rigorous inspection of solutions (\ref{solution}-\ref{solution2}) shows that the only difference between $\varepsilon_{j,sh}$ and $\varepsilon_{j,in}$ is the presence of the term $(-C_{sh} \kappa'_{in})$ in the former and $(-C_{in} \kappa'_{sh})$ in the latter expression. For any internal structure model and subscript $j$ considered, the numerical values of these terms are of the same sign (negative) as the numerical value of the sum of the other terms of the numerator. Since the moment of inertia $C_{in}$ is one order of magnitude larger than $C_{sh}$, while $\kappa'_{sh}$ and $\kappa'_{in}$ are of the same order of magnitude, we have $|\varepsilon_{j,in}|>|\varepsilon_{j,sh}|$. The difference between $\varepsilon_{j,sh}$ and $\varepsilon_{j,in}$ is relatively small (less than one order of magnitude) since $(-C_{in} \kappa'_{sh})$ is of the same order of magnitude or slightly smaller than the other terms in the numerator and since $(-C_{sh} \kappa'_{in})$ is one to two orders of magnitude smaller than the other terms in the numerator.

Time evolution of the shell obliquity $\varepsilon_{sh}$, compared to the solid case obliquity $\varepsilon$, is given in Fig. 6 for the internal structure models with the smallest ($0.033^\circ$) and the highest ($0.044^\circ$) value of $\varepsilon_{1,sh}$. Time variations are of the order of 6 percent in both the ocean case and the solid case, because of the similarly small influence of the terms $j>1$ for both cases. The ocean could be detected if the measured obliquity is different from the obliquity of the solid case. Since the phases $\gamma_j$ are the same for both the solid and the ocean cases, the time-variations are in phase for the three curves, and we conclude that the measurement error $\sigma$ should be smaller than $|\varepsilon_{1}-\varepsilon_{1,sh}|/3$, for the difference to be at $3\sigma$, which is in the range $[0.0037^\circ,0.0073^\circ]$, corresponding to a surface displacement of about $100$m-$200$m. Such an accuracy will be difficult to achieve with forthcoming ground-based radar measurements proposed in Margot (2011). Future space missions could provide a sufficient precise obliquity value by imaging (see Stiles et al. 2008 and 2010 for Titan and Pfyffer et al. 2011 for Mercury). As the detection of the ocean is already challenging, constraining the internal parameters, such as the shell thickness $h_{sh}$, will be more complicated or even impossible. Indeed, the dependence of the obliquity amplitudes with respect to $h_{sh}$ is not as clear as was the dependence of $\varepsilon_{j,m}$ with respect to $h_m$ for Io (see Fig. 5). This is due to the non-unicity of the solution for the internal structure when $h_{sh}$ is fixed and to the fact that the obliquity amplitudes depend significantly on the other internal parameters.

\subsection{Ganymede}

Like Europa, Ganymede probably has a liquid water ocean ($o$) beneath a water ice shell ($sh$) (Kivelson et al. 2002). Besides, this satellite has a more complex internal structure since it is the most differentiated Galilean satellite (Anderson et al. 1996). Unlike Europa, an ice mantle ($i$) isolates the liquid water ocean ($o$) from the rock mantle ($r$). In addition, Ganymede is thought to have a fluid core ($f$) over its solid metallic core ($c$), since an intrinsic magnetic field has been detected (see Kivelson et al. 1998). A range of solutions for the internal structure is defined by considering equally-spaced values for the shell thickness $h_{sh}$, the radii $R_i$, $R_r$, $R_f$, and $R_c$, and the densities $\rho_{sh}$, $\rho_o$, $\rho_i$, and $\rho_c$, which can be found in Baland and Van Hoolst (2010). We then compute $\rho_{r}$ and $\rho_f$ from the mass $M$ and the mean moment of inertia $I$. The retained solutions satisfy the conditions $3000\leq\rho_r\leq3800$ kg/m$^3$, $5150\leq\rho_f\leq 8000$ kg/m$^3$, $R_o>R_i>R_r>R_f>R_c$ and $\rho_{sh}\leq\rho_{o}\leq\rho_{i}$. We consider that the interior ($in$) is made of the ice and rock mantles and of the fluid and solid cores, because the state of core should not have a noticeable influence on the rotation state of the satellite, in the same way that it does not play a significant role in the librations, because of the small size and of the small flattening of the core (Baland and Van Hoolst 2010).

As expected, the free period of the solid case ($T_f=19.97$ years) is in the range of values of the 'coupled' free period $T_+$  ($17$-$42$ years, see Fig. 3) computed from Eq. (\ref{omegapm}). As for Europa, the obliquity amplitudes for $j=1, 2, 4, 6$ or $7$ correspond to periods $T_j$ far from the opposite of the free periods $T_+$ and $T_-$ and an order of magnitude analysis of the terms of the numerators in Eqs. (\ref{solution}-\ref{solution2}) leads also to the conclusion that $\varepsilon_{j,sh}$ and $\varepsilon_{j,in}$ differ from each other by less than one order of magnitude and that $|\varepsilon_{j,in}|>|\varepsilon_{j,sh}|$ (see Fig. 4). The obliquity amplitudes for $j=3,5$ or $8$ do not follow these conclusions. An order of magnitude analysis for $j=3$ shows that $\varepsilon_{3,sh}$ and $\varepsilon_{3,in}$ also differ slightly from each other, but since $-T_3=30.22$ years is in the range of $T_+$, the denominator can be so small that both $\varepsilon_{3,sh}$ and $\varepsilon_{3,in}$ can be much larger than in the solid case, and of the same order of magnitude or larger than $\varepsilon_{1,sh}$ and $\varepsilon_{1,in}$, for some specific internal structure models.
The period $-T_5=11.86$ years is far enough from $T_+$ and $T_-$ to avoid such a significant resonant amplification. For $j=5$, the term $(-C_{sh} \kappa'_{in})$ can be as large as the other part of the numerator and of a different sign, leading to very small values of $\varepsilon_{5,sh}$. The term $(-C_{in} \kappa'_{sh})$ is always at least as large as the other terms of the numerator and of the same sign, and $\varepsilon_{5,in}$ is quite close to $\varepsilon_{5}$. The range of values for $T_-$ goes from $0.04$ to $9.5$ years and contains $-T_8=0.68$ years. For $j=8$, the term $(-C_{sh} \kappa'_{in})$ is two orders of magnitude smaller than the other terms of the numerator, while $(-C_{in} \kappa'_{sh})$ can be of the same order of magnitude and of a different sign than the other part of the numerator. As a result, $\varepsilon_{8,sh}$ cannot be significantly smaller than in the solid case, while $\varepsilon_{8,in}$ might be very small. For a small fraction of the internal structure models considered, a resonant amplification allows $\varepsilon_{8,sh}$ to compete with $\varepsilon_{1,sh}$  while $\varepsilon_{8,in}$ is also amplified but not enough to reach the level of $\varepsilon_{1,in}$ (Fig. 4). 

We conclude that only the shell obliquity amplitudes $\varepsilon_{1,sh}$, $\varepsilon_{3,sh}$ and $\varepsilon_{8,sh}$ can play a significant part in the actual value of the time dependent shell obliquity $\varepsilon_{sh}$. Their range of possible values are given in Table 6. The first obliquity amplitude presents a similar behavior with respect to the ice shell thickness as in the case of Europa, that is to say that it increases with increasing $h_{sh}$ but also significantly depends on the other internal structure parameters (Fig. 5). The amplitudes $\varepsilon_{3,sh}$ and $\varepsilon_{8,sh}$ behave differently because of the significant resonant amplification that occurs for some specific internal structure models. A large value of $\varepsilon_{3,sh}$ or $\varepsilon_{8,sh}$ is associated with a thin ice shell (Fig. 5), but not all models with a thin shell yield large values $\varepsilon_{3,sh}$ or $\varepsilon_{8,sh}$. Because of these resonances, the shell obliquity behavior over time depends strongly on the chosen internal structure model. In Fig. 6, the solid case is compared to the ocean case for three specific internal structure models. The first two models correspond respectively to the minimal and maximal values of $\varepsilon_{1,sh}$. The third model presents a moderate resonant amplification of $\varepsilon_{8,sh}$. All models present a significant resonant amplification of $\varepsilon_{3,sh}$, the amplification being largest for the model corresponding to the maximal value of $\varepsilon_{1,sh}$.

The ocean could be detected quite easily if the actual internal structure corresponds to a shell obliquity behavior very different from the solid case at the time of measurement. If $\varepsilon_{8,sh}$ is not strongly amplified, the best measuring times are those at which the different terms of the solution combine to provide the maximal obliquity, that is to say around 2030, 2070, 2110 and so on. As an example, if the measurement takes place around the year 2030, and if the actual internal structure is close to the one corresponding to the minimal value of $\varepsilon_{1,sh}$, an accuracy of $0.015^\circ$ corresponding to a surface displacement of about $700$ m would be sufficient. The future space mission JUICE (JUpiter ICy moons Explorer, under study by ESA for the first L-class mission) provides a sufficient accuracy. However, constraining the internal structure will be difficult, since many very different internal structure models could fit to a same observation, through different combinations of the three main obliquity amplitudes. For example, Fig. 6 shows that the solid case model, the model with the minimal value of $\varepsilon_{1,sh}$ which has a shell thickness $h_{sh}=25$ km, and the model presenting a moderate resonant amplification of $\varepsilon_{8,sh}$ and such as $h_{sh}=37.5$ km, have similar shell obliquity values around the years 2013-2014. This also means that observing a shell obliquity consistent with the solid case does not necesseraly means that there is no ocean. Obliquity measurements spread over several decades could help determine the individual obliquity amplitudes and if Ganymede has a rather thin ice shell or not.

\subsection{Callisto}

Callisto is similar to Ganymede in size and mean density, but its mean moment of inertia is significantly larger. Therefore, Callisto is less differentiated than Ganymede. Following Anderson et al. (2001a) and Khurana et al. (1998), we consider that Callisto is made of a water ice shell (sh), a liquid water ocean (o) and an interior (in) made of a homogeneous mixture of ice, rock and iron. The internal structure models considered in the following are defined by choosing equally-spaced values for the thickness of the ice shell $h_{sh}$ and for the densities of the shell $\rho_{sh}$ and of the ocean $\rho_o$ and by computing the radius $R_{in}$ and the density $\rho_{in}$ of the interior such that the model satisfies the mass $M$ and the mean moment of inertia $I$ (Baland and Van Hoolst, 2010). The retained solutions are such that $\rho_o\geq\rho_s$ and that the radius of the ocean $R_o=R-h_{sh}>R_{in}$.

For this range of internal structure models, the free period $T_+$ is between $181.7$ and $233.7$ years (Fig. 3). This range contains the free mode period of the solid case ($T_f=203.1$ years) and is close enough to the period $-T_{2}=137.7$ years to allow $\varepsilon_{2,sh}$ to be non-negligible compared to $\varepsilon_{1,sh}$, in the same way as $\varepsilon_{2}$ was not negligible compared to $\varepsilon_{1}$ in the entirely solid satellite case. The range of possible values for $T_-$ is $[3.97,58.42]$ years and contains $-T_{4}=11.86$ years and $-T_{5}=30.22$ years, which means that some specific internal structure models present a resonant amplification such that the obliquity amplitudes $\varepsilon_{4,sh}$ or $\varepsilon_{5,sh}$ are amplified to the level of $\varepsilon_{1,sh}$ (Fig. 4). For $j\neq 4$ or $5$, the obliquity amplitudes $\varepsilon_{j,sh}$ and $\varepsilon_{j,in}$ differ from each other by less than one order of magnitude and $|\varepsilon_{j,in}|>|\varepsilon_{j,sh}|$ (Fig. 4), similarly as for the obliquity amplitudes of Europa and for the obliquity amplitudes of Ganymede with $j\neq3,5$ or $8$. For $j=4$ or $j=5$, the obliquity amplitudes can be much smaller as well as much larger than the entirely solid case values, as a result of the negative frequencies $\dot \Omega_{j}$ being quite large. For some internal structure models, the common part of the numerator of solutions (\ref{solution}-\ref{solution2}) can be positive and of the same order of magnitude as the terms $(-C_{sh} \kappa'_{in})$ and $(-C_{in} \kappa'_{sh})$ which are negative. As a result, $\varepsilon_{4/5,sh/in}$ can be very small. For other internal structure models, the order of magnitude analysis shows that $\varepsilon_{4/5,sh/in}$ behave as for $j\neq 4$ or $5$, unless they are amplified by a resonance due to the closeness of $-T_{4/5}$ with respect to $T_-$. 

The first obliquity amplitude behaves with respect to the ice shell thickness as the first obliquity amplitude of Europa and Ganymede (Fig. 5). It increases slightly with increasing $h_{sh}$ but depends also on the other characteristics of the internal structure. The amplitudes $\varepsilon_{2,sh}$, $\varepsilon_{4,sh}$ and $\varepsilon_{5,sh}$ have a different behavior, because of the significant resonant amplification that occurs for some internal structure models. All models present a moderate amplification of $\varepsilon_{2,sh}$ while a large amplification of $\varepsilon_{4,sh}$ or $\varepsilon_{5,sh}$ occurs only for shell thicknesses of about $100$ and $200$ km, respectively. The range of possible values for the dominating obliquity ampltitudes are given in Table 6.

The time variable shell obliquity for the interior structure model corresponding to the minimum and maximum values of $\varepsilon_{1,sh}$ ($0.053^\circ$ and $0.133^\circ$) is presented in Fig. 6. Both internal structure models are characterized by a significant resonant amplification of the second term of the solution and have obliquity amplitudes $\varepsilon_{2,sh}$ close to $0.05^\circ$. As a result, the 'minimal model' can reach a zero shell obliquity. Fig. 6 also shows the evolution of the shell obliquity for specific interior structure models which present a resonant amplification of the fourth or the fifth term of the solution, illustrating that a lot of internal structure models can have the same shell obliquity at a given time. Regarding the possibility of constraining the internal structure from a shell obliquity measurement at a given time, the problem is similar as for Ganymede. Two or even three dominant terms of the solution combine to give the main part of the solution, and each of the dominant obliquity amplitudes does not depend uniquely on any parameter of the internal structure. A single measurement would only allow detecting the internal ocean, provided that the measured obliquity is sufficiently different from the solid case obliquity. If $\varepsilon_{4,sh}$ and $\varepsilon_{5,sh}$ are not strongly amplified, the largest difference in the time interval (2010-2040) is about $0.13^\circ$. Again assuming a $3\sigma$ error as large as this difference, the measurement precision of the rotation axis on Callisto's surface should be about $2$ km, which is possible wih a mission like JUICE. As for Ganymede, a measured obliquity compatible with the solid case does not mean that there is no internal ocean, since several internal structure models with an ocean can have, at a given time, a similar obliquity as in the rigid case. A long-term monitoring of the shell obliquity of Callisto would be more helpful to discrimate among the possible internal structure models.

\section{Discussion and conclusion}

We have computed the obliquity of the four Galilean satellites for a wide and representative range of internal structure models with homogeneous layers, consistent with the constraints on radius, mass and mean moment of inertia provided by the Galileo spacecraft mission. We have assumed that the satellites are locked in a Cassini state. Therefore, we used a model for the obliquity based on the Cassini state and extended to the presence of an internal liquid layer, including the internal gravitational and pressure torques induced by the presence of this layer. The model used in this study is also generalised to a multi-frequency orbital precession of the satellites. With this model, we have shown that the surface layer obliquity of Io does not present a large variety for the different internal structure models considered. The obliquity of Europa is almost constant and its mean value is larger for a thicker ice shell, although it depends also on other characteristic of its interior. The internal structure models of Ganymede and Callisto present more various behaviors in the shell obliquity, because of the possibility of combined resonant amplifications for some periodic terms of the solution.

We have compared the time-behavior of the surface layer obliquity with the behavior in the case of an entirely solid satellite, to assess the possibility to detect the considered liquid layer from a single or multiple obliquity measurement. For Io and Europa, an accuracy of about $0.16$ m and $150$ m would be needed, respectively, on the location of the rotation axis at the surface. The ocean of Ganymede and Callisto may be more easily detected if the actual internal structure model would lead to a very different value of the obliquity than in the solide case. In the best case, a precision of about $700$ m and $2$ km for Ganymede and Callisto would be sufficient. However, for Ganymede and Callisto, a measured obliquity compatible with the entirely solid case would not exclude the existence of an internal ocean, since some internal structure models with an ocean cannot be distinguished from the solid case at some given times. In any case, for Ganymede and Callisto, the combination of many obliquity measurements spread over several decades would be needed to learn more about their internal structure. This is a reason why the planned ground observations of the rotation state of Ganymede is of interest, though being probably less accurate than any potential measurements from a spacecraft that would reach Ganymede around the year 2030. To interpret forthcoming measurements, one has to ensure that the error on the knowledge of the orbital plane orientation of the considered satellite is properly taken into account, since the rotation axis orientation would be measured with respect to a standard reference frame which is the ICRF, and not directly with respect to the orbital plane. 

We note that the theoretical values for the obliquity of Europa with (of the order of $0.04$ deg) or without ocean ($0.055$ deg) are not consistent with the estimates of about $1$ deg by Rhoden et al. (2010, 2011), inferred from the observations of cycloidal lineaments and strike-slip fault patterns. The pressure effect on the density profile cannot be of a great importance because of the relative smallness of Europa and cannot explain the mismatch. Also the assumption of a different density profile not associated with the normalized mean moment of inertia of 0.356 related to the assumption of hydrostatic equilibrium leads only to a difference of about 10\% on Europa's obliquity with respect to our results. This indicates that it is impossible with our obliquity model and for any reasonable internal structure model for Europa to reach the high obliquity value of Rhoden et al. (2010, 2011), unless a free mode would be excited to such a value because of a recent excitation. The obliquity of Rhoden's studies may reflect the past rotation state of Europa associated with the past orbital state and its past internal structure. In our study, we use the ephemerides of Lainey et al. (2006) which are valid for a short time span in the past (of the order of $10^3$ years) compared to the age of Europa's surface ($10^8$ years) and we are of course considering the present internal structure.\\

We have considered internal structure models for the satellites that have homogeneous and uniform layers with flatennings computed by solving Clairaut's equation, in order to keep a relative simplicity in our assessment of the effect of interior parameters such as the ice shell thickness on the solution for the rotation axis orientation. However, any forthcoming measurement would need to be interpreted in terms of more realistic models for the internal structure, in particular if the measurement indicates that a resonant amplification is at work. Indeed, in a resonant case, the solution is very sensitive to any change in the internal structure. For instance, the pressure effect on the density of the materials could be considered for Ganymede and Callisto which are large satellites. Another possibility would be that the icy satellites slightly depart from the hydrostatic equilibrium because, for instance, of differential tidal heating which could flatten the poles (lateral shell thickness variations), as it is assumed to be the case for Titan (Nimmo and Bills 2010). We note that the general solution for the Cassini state that we have derived in this paper is valid also for non-hydrostatic internal structure models and or non hydrostatic flattenings, allowing us to realize some tests. For instance, multiplying and dividing the polar flatennings of the surface and of the ocean-ice interface, respectively, by a factor two with respect to the hydrostatic case decreases the first obliquity amplitude of Europa by only about $15\%$. Given the unknown internal structure of the Galilean satellites, it seems reasonable for now to consider hydrostatic flattenings as a good first order approximation. Forthcoming measurements would help to test and improve our model.

\section*{Acknowledgments}
We thank two anonymous reviewers for their comments which helped to improve the paper. 
R.M.B. is funded by the Conseil R\'egional des Pays de la Loire, France. Part of this work was carried out at the Royal Observatory of Belgium, under the funding of the Fonds pour la formation \`{a} la Recherche dans l'Industrie et dans l'Agriculture (FRIA). This work was financially supported by the Belgian PRODEX program managed by the European Space Agency in collaboration with the belgian Federal Science Policy Office.

\appendix

\section{Angular momentum equation}
\label{Append}

We here provide a concise demonstration of Eq. (\ref{rigidcase}). The angular momentum equation is written in an inertial frame, which is here taken to be a reference frame attached to the Laplace plane and centered at the center of mass of the satellite, as
 \begin{equation}\label{A11}
  \frac{d \vec L_{IN}}{dt}=\vec\Gamma_{IN}
 \end{equation}
with $\vec L_{IN}$ the angular momentum of the satellite and $\vec\Gamma_{IN}$ the torque exerted on the satellite by the planet. The expression of the torque in the frame attached to the principal axes of inertia of the satellite (Body Frame) is (Murray and Dermott 1999, Eqs 5.43-5.45):
\begin{equation}
\vec \Gamma_{BF}=\left(
\begin{array}{c}
 3 n^2 a^3 (C-B) Y Z /d^5 \\
 3 n^2 a^3 (A-C) Z X /d^5 \\
 3 n^2 a^3 (B-A) X Y /d^5
\end{array}\right)
\end{equation}
with $A<B<C$ the principal moments of inertia of the satellite, $n$ its mean motion, $a$ its semi-major axis, $d$ the distance between the satellite and the planet, and $(X,Y,Z)$ the position of the planet in the Body Frame in cartesian coordinates. If we note $(\psi,\theta,\phi)$ the Euler angles orienting the Body Frame with respect to the inertial frame (see Fig. 7), the position of the planet is 
\begin{equation} 
\left(
\begin{array}{c}
 X \\
 Y \\
 Z
\end{array}\right)=R_z(\phi).R_x(\theta).R_z(\psi-\Omega).R_x(-i).R_z(-(\omega-\pi)-f)\left(
\begin{array}{c}
d \\
 0 \\
 0
\end{array}\right)
\end{equation}
with $\Omega$, $i$, $\omega$, and $f$ the node longitude, inclination, pericenter longitude, and true anomaly of the satellite's orbit with respect to the Laplace plane (the pericenter of the planet seen as in orbit around the satellite is then $\omega-\pi$), and the rotation matrices defined as
\begin{eqnarray}\label{RxRz}
R_x[\theta] = \left(
\begin{array}{ccc}
 1 & 0 & 0 \\
 0 & \cos{\theta} & \sin{\theta} \\
 0 & -\sin{\theta} & \cos{\theta}
\end{array}
\right) \qquad R_z[\theta] = \left(
\begin{array}{ccc}
 \cos{\theta} & \sin{\theta} & 0 \\
 -\sin{\theta} & \cos{\theta} & 0 \\
 0 & 0 & 1
\end{array}
\right).
\end{eqnarray}

Correct up to the first order in orbital eccentricity $e$,
\begin{eqnarray}
f&\simeq&M+2e \sin{M}\\
d&\simeq&a-a e \cos{M}
\end{eqnarray}
with $M$ the mean anomaly. Because of the synchronicity of the rotation with the orbital revolution, the Euler angle $\phi\simeq -\psi+\Omega+\omega-\pi+M$ (e.g. Peale 1969), correct up to the first order in the small angles $i$ and $\theta$, and neglecting librations (the variations of the rotation rate). Therefore, at the first order in $i$, $\theta$ and $e$, we have that 
\begin{eqnarray}\label{A7}
\vec \Gamma_{BF}\simeq\left(
\begin{array}{c}
 0 \\
 3 n^2 (C-A) [i \sin{(\omega+M)}-\theta \sin{(\omega+M+\Omega-\psi)}] \\
 6 n^2 (B-A) e \sin M
\end{array}\right).
\end{eqnarray}

The torque is then expressed in the inertial frame thanks to the appropriate rotations
\begin{equation}
 \vec \Gamma_{IN}=R_z(-\psi).R_x(-\theta).R_z(-\phi).\vec \Gamma_{BF}.
\end{equation}
Finaly, the torque is averaged over an orbit period with the slowly varying $\omega$, $\Omega$, and $\psi$ held constant, and since 
\begin{equation}
 (s_x,s_y)\simeq(\theta\cos{(\psi-\pi/2)},\theta\sin{(\psi-\pi/2)})
\end{equation} and
\begin{equation}
(n_x,n_y)\simeq(i\cos{(\Omega-\pi/2)},i\sin{(\Omega-\pi/2)}),
\end{equation}
the torque becomes
\begin{eqnarray}\label{TIN}
\vec \Gamma_{IN}\simeq\left(
\begin{array}{c}
 \frac{3}{2} n^2 (C-A) (i \cos{\Omega}-\theta\cos{\psi})\\
 \frac{3}{2} n^2 (C-A) (i \sin{\Omega}-\theta\sin{\psi}) \\
 0
\end{array}\right)
=\frac{3}{2} n^2 (C-A) (\hat s \wedge \hat n),
\end{eqnarray}
and depends on only one moment of inertia difference, namely $(C-A)$. The physical reason for only having $(C-A)$ in the torque 
expression is related to the fact that the $B$-axis is on average perpendicular to the satellite-planet axis. \\%, as might be anticipated from Eq. (\ref{A7})\\ 

Neglecting wobble, the angular momentum in the Body Frame of the satellite is given by
\begin{equation}
\vec L_{BF}=\left(
\begin{array}{c}
 0 \\
 0 \\
 C (\dot\phi+\dot\psi \cos{\theta}) 
\end{array}\right)
\end{equation}
and can be expressed, at first order in $\theta$ and with $(\dot\phi+\dot\psi\simeq n)$, in the inertial frame as
\begin{equation}\label{LIN}
\vec L_{IN}\simeq\left(
\begin{array}{c}
 n\, C\, \theta\sin\psi\,  \\
 -n\, C\, \theta\cos\psi\,  \\
 n\, C 
\end{array}\right)\simeq n\, C\, \hat s.
\end{equation}

With the inclusion of Eqs (\ref{TIN}) and (\ref{LIN}), angular momentum equation (\ref{A11}) becomes 
 \begin{equation}\label{AFinal}
  n\, C \frac{d \hat s}{dt}=\frac{3}{2} n^2 (C-A) (\hat s \wedge \hat n)
 \end{equation}
which is equivalent to Eq. (\ref{rigidcase}). The interested reader is invited to consult Appendix B of Noyelles (2010) who also find that the torque is proportional to $(C-A)$. We note that equation (\ref{AFinal}) is consistent with the small angle approximation which can be obtained from Eq. (1) of Ward (1975). Although Ward (1975) is citing Peale (1969), the corresponding equation of Peale (1969) is not consistent with the one of Ward (1975) and our Eq. (\ref{AFinal}) because of a factor 2 error in Peale (1969) which has been recognized in Peale (1988).

\newpage

\noindent{\bf TABLES}

\begin{table}[htdp]
\begin{center}
\begin{tabular}{lccccccc}
\hline
& $R$ [km] & $\bar\rho$ [kg\, m$^{-3}$] &$C_{20}$ & $C_{22}$ & $\frac{I(\simeq C)}{MR^2}$  &$n$ (rad/day)&  $T_f$ (years)\\
\hline
Io 	 & $1821.6$ & $3527.5$ &$1859.5\times 10^{-6}$& $558.8\times 10^{-6}$ & $0.378$ & $3.551552286182 $ & $0.410$\\ %255$\\ %0.596256$\\
Eur.   & $1565.0$ & $2989.0$ &$435.5\times 10^{-6}$ & $131.5\times 10^{-6}$ & $0.346$ & $1.769322711123 $ & $3.211$\\ %71$\\%$3.21762$\\
Gan.& $2631.2$ & $1942.0$ &$127.53\times 10^{-6}$& $38.26\times 10^{-6}$ & $0.312$ & $0.878207923589 $ & $19.967$\\ %3$ \\
Cal. & $2410.3$ & $1834.4$ &$ 32.7\times 10^{-6}$ & $ 10.2\times 10^{-6}$ & $0.354$ & $0.376486233434 $ & $203.076$ \\
\hline
\end{tabular}
\end{center}
\caption{\label{Tab1}Mean radius, mean density, second-degree gravity field coefficients, mean moment of inertia (according to Schubert et al. 2004), mean motion (Lainey et al. 2006) and free period of precession for a solid satellite (using Eq. \ref{wf} of this paper).} 
\end{table}

\begin{table}[htdp]
\begin{center}      
\begin{tabular}{c c c c c c }   
\hline   $j$&$i_j$&$\dot\Omega_j$&period&$\gamma_j$&$\varepsilon_j$  \\
  &(deg)&(rad/year)& (years)&(deg)&(deg)\\    
\hline                    
      $1$&$0.0360066 $&$-0.845589 $&$-7.43054 $&$ 21.8492$&$0.0021042 $  \\
      $2$&$ 0.0103611 $&$-0.207903 $&$-30.2218 $&$-78.3448 $&$0.0001426 $  \\
      $3$&$ 0.00188438 $&$-0.0456245 $&$-137.715 $&$52.8155 $&$ 0.00000563$  \\
      $4$&$0.000635096 $&$-0.0111625 $&$-562.884 $&$203.798 $&$ 0.000000463$  \\
      $5$&$ 0.000410817 $&$-9.22056 $&$ -0.681432 $&$-89.7306 $&$ 0.000621513 $  \\
      $6$&$0.000277227 $&$1.05936 $&$ 5.93112 $&$180.212 $&$ -0.000017935$  \\
\hline
\end{tabular}
\end{center}
\caption{\label{Tab2}Columns 2 to 5: amplitudes, frequencies, periods, and phases of the orbital precession of Io adapted from Lainey et al. (2006). They give the orbital precession in the equatorial plane of Jupiter and used J1950 as the time origin. Here we consider the Laplace plane and we keep the J1950 time origin. The Laplace plane has a node of $138.37395^\circ$ and an inclination (also called tilt) of $0.00194004^\circ$ with respect to the equatorial plane of Jupiter. Since the tilt is small, the frequencies and amplitudes of the orbital precession are almost the same with respect to the Laplace plane as to the equatorial plane of Jupiter. The $x$-axis of the Laplace plane is taken as the node of the Laplace Plane on the equatorial plane of Jupiter. The obliquity amplitudes of the solid Cassini state model are given in the last column.} 
\end{table}

\begin{table}[htdp]
\begin{center}       
\begin{tabular}{c c c c c c  }   
\hline   $j$&$i_j$&$\dot\Omega_j$&period&$\gamma_j$&$\varepsilon_j$   \\
  &(deg)&(rad/year)& (years)&(deg)&(deg)\\    
\hline                    
      $1$&$0.463007$	&$-0.207903 $	&$-30.2218 $	&$-78.2534$	&$0.055036$	  \\
      $2$&$0.025215$	&$-0.0456245 $	&$-137.175 $	&$52.9069$	&$0.000602$	  \\
      $3$&$0.00676408$	&$-0.0111626 $	&$-562.877 $	&$203.887 $	&$0.00003880$ \\
      $4$&$-0.00120351$	&$-0.845589 $	&$-7.43054 $	&$21.945 $	&$-0.00091571$  \\
      $5$&$-0.00117961$	&$-9.22056 $	&$-0.681432 $	&$-89.6391 $	&$0.00149742$	 \\
      $6$&$0.000831874$ &$1.05936 $	&$5.93112 $	&$ 180.303 $	&$-0.000292163$  \\
      $7$&$0.000210828$ &$-0.529634 $	&$-11.8633 $	&$-121.964 $	&$ 0.000078232$  \\
\hline
\end{tabular}
\end{center}
\caption{\label{Tab3}Columns 2 to 5: amplitudes, frequencies, periods, and phases of the orbital precession of Europa adapted from Lainey et al. (2006). As for Io, we express the orbital precession with respect to the Laplace plane. The Laplace plane has a node of $138.28257^\circ$ and an inclination (also called tilt) of $0.0190513^\circ$ with respect to the equatorial plane of Jupiter. The obliquity amplitudes of the solid Cassini state model are given in the last column.} 
\end{table}

\begin{table}[htdp]
\begin{center}      
\begin{tabular}{c c c c c c  }   
\hline   $j$&$i_j$&$\dot\Omega_j$&period&$\gamma_j$&$\varepsilon_j$   \\
  &(deg)&(rad/year)& (years)&(deg)&(deg)\\    
\hline                    
      $1$&$0.182576$&$-0.0456245$&$-137.715$&$52.9125$&$0.030961$  \\
      $2$&$0.04026$&$-0.0111624$&$-562.888$&$203.898$&$0.001481$  \\
      $3$&$-0.0165233$&$-0.207903$&$-30.2218$&$-78.2478$&$-0.0321738$  \\
      $4$&$0.00180254$&$1.05936$&$5.93112$&$180.311$&$-0.00138973$  \\
      $5$&$0.000288346$&$-0.529633$&$-11.8633$&$-121.96$&$-0.000710451$  \\
      $6$&$0.000234188$&$0.529663$&$11.8626$&$-37.1349$&$ -0.000146909$  \\
      $7$&$0.00020561$&$1.58901$&$3.95416$&$122.814$&$-0.000171623$  \\
      $8$&$0.000156054$&$-9.22056$&$-0.681432$&$-89.6335$&$ -0.000161568$  \\
\hline
\end{tabular}
\end{center}
\caption{\label{Tab4}Columns 2 to 5: amplitudes, frequencies, periods, and phases of the orbit precession of Ganymede adapted from Lainey et al. (2006). We express the orbital precession with respect to the Laplace plane. The Laplace plane has a node of $138.27696^\circ$ and an inclination (also called tilt) of $0.0977821^\circ$ with respect to the equatorial plane of Jupiter. The obliquity amplitudes of the solid Cassini state model are given in the last column.} 
\end{table}

\begin{table}[htdp]
\begin{center}       
\begin{tabular}{c c c c c c  }   
\hline  $j$&$i_j$&$\dot\Omega_j$&period&$\gamma_j$&$\varepsilon_j$   \\
  &(deg)&(rad/year)& (years)&(deg)&(deg)\\    
\hline                    
      $1$&$0.257303 $&$-0.0111625 $&$-562.883 $&$203.903 $&$0.145222$  \\
      $2$&$-0.0298451 $&$-0.0456245 $&$-137.715 $&$52.9117 $&$0.0927289$  \\
      $3$&$0.00380572 $&$1.05936 $&$5.9311 $&$180.311 $&$-0.00369772$ \\
      $4$&$0.000569814 $&$-0.529633 $&$-11.8633 $&$-121.955 $&$-0.000605166$  \\
      $5$&$-0.000566284 $&$-0.207901 $&$-30.2219 $&$-78.2489 $&$0.000665294$  \\
      $6$&$0.000503546 $&$0.529662 $&$11.8626 $&$-36.9538 $&$-0.000475755$  \\
      $7$&$0.000431194 $&$1.58901 $&$3.95416 $&$122.806 $&$-0.000422958$ \\
      $8$&$-0.000353182 $&$1.07066 $&$5.86852 $&$-23.145 $&$0.000343263$  \\
\hline
\end{tabular}
\end{center}
\caption{\label{Tab5}Columns 2 to 5: amplitudes, frequencies, periods, and phases of the orbit precession of Callisto adapted from Lainey et al. (2006). We express the orbital precession with respect to the Laplace plane. The Laplace plane has a node of $138.27718^\circ$ and an inclination (also called tilt) of $0.440296^\circ$ with respect to the equatorial plane of Jupiter. The obliquity amplitudes of the solid Cassini state model are given in the last column.} 
\end{table}

\begin{table}[htdp]
\begin{center}       
\begin{tabular}{l l}   
\hline  Satellite& Range of the dominating obliquity amplitudes of the surface layer (deg) \\
\hline  Io& $\varepsilon_{1,m}\in[0.00198,0.00208]$ and $\varepsilon_{5,m}\in[0.00054,0.00060]$  \\
Europa& $\varepsilon_{1,sh}\in[0.0330,0.0444]$  \\
Ganymede& $\varepsilon_{1,sh}\in[0.0085,0.0320]$, $\mid\varepsilon_{3,sh}\mid\in[0.0068,\textrm{a few degrees}]$,  \\    
               & and $\varepsilon_{8,sh}\in[-0.0193,0.0067]$  \\
Callisto& $\varepsilon_{1,sh}\in[0.0534,0.1325]$, $\varepsilon_{2,sh}\in[0.0451,0.0584]$,  \\
               & $\varepsilon_{4,sh}\in[-0.0073,0.0616]$ and $\varepsilon_{5,sh}\in[-0.6021,0.0530]$  \\
\hline
\end{tabular}
\end{center}
\caption{\label{Tab6}Range of the dominating obliquity amplitudes for the mantle of Io and the ice shell of Europa, Ganymede, and Callisto, corresponding to the range of internal structure models considered (see also Fig. 5). For Ganymede, a resonant amplification can lead to values for $\varepsilon_{3,sh}$ of the order of several tens of degrees, but realistic values cannot exceed a few degrees, since the obliquity of the Galilean satellites is thought to be small.} 
\end{table}

 \newpage
 \noindent{\bf FIGURES}
 \vspace{1cm}

\begin{figure}[!htb]
\begin{center}
\includegraphics[width=14cm]{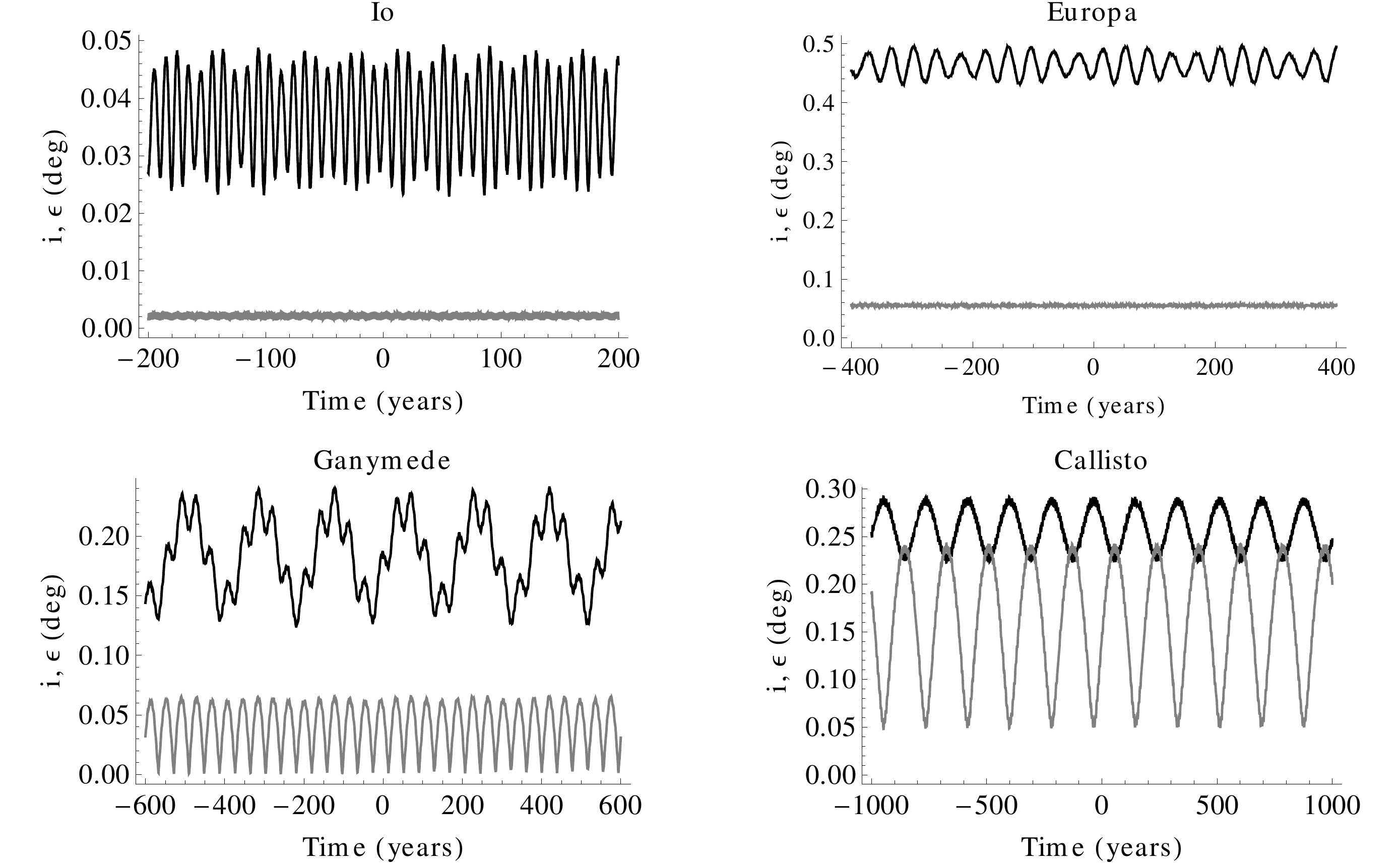}\\
 %(centre en 1950, mais apparemment Bills centre en 1980...)
\caption{Time evolution of the orbital inclination of the four Galilean satellites with respect to their respective Laplace plane (black) and time evolution of their obliquity in the case where these satellites are entirely solid (gray). We have chosen J2000 as the time origin. The chosen time spans have the same widths as in Bills (2005), in order to compare with its third figure.}
\end{center}
\end{figure}

% \begin{figure}[!htb]
% \begin{center}
% \hspace{0cm}
% \includegraphics[width=15cm]{solid_case_compbills.png}\\
% Fig. 2 
% \end{center}
% \end{figure}

\begin{figure}[!htb]
\begin{center}
\includegraphics[width=6cm]{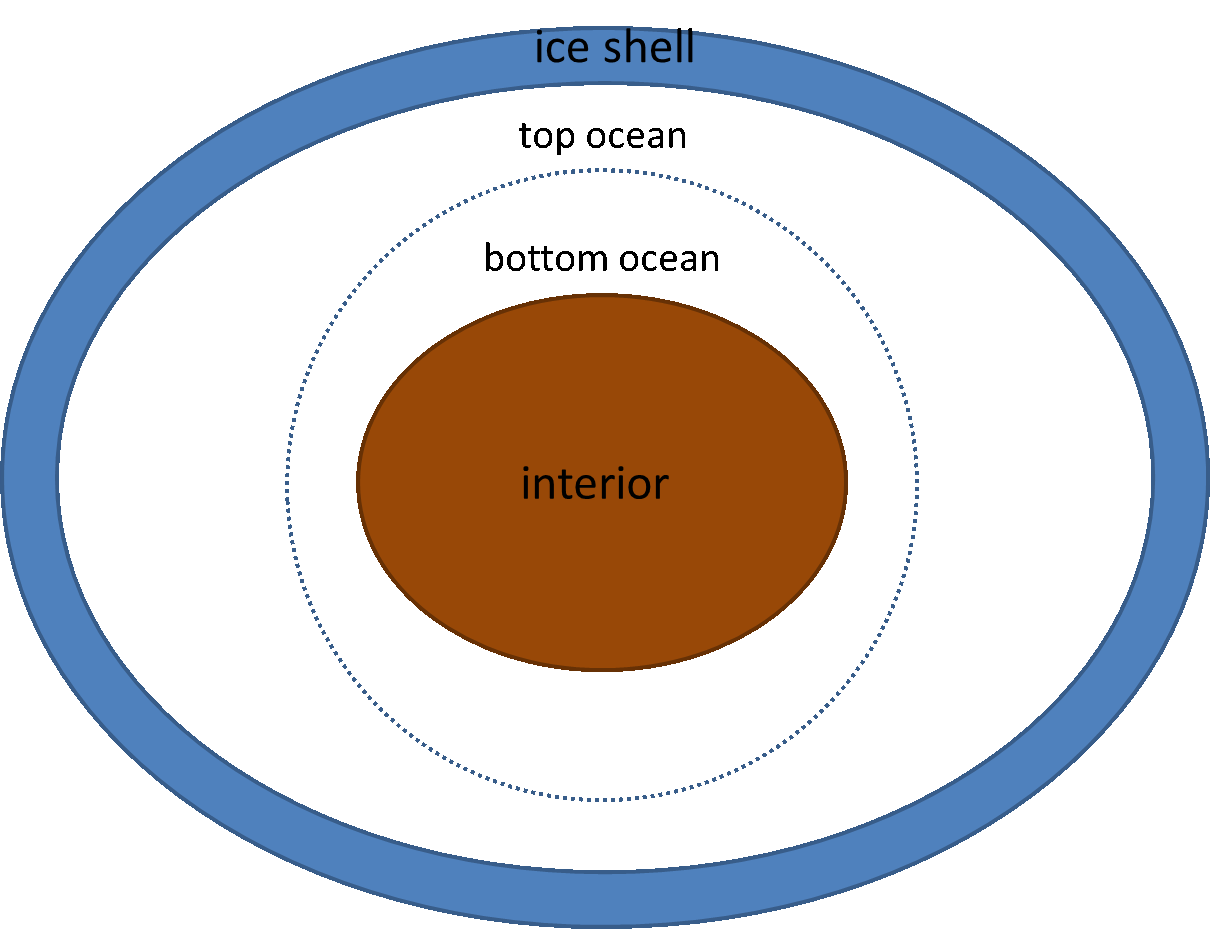}\qquad \includegraphics[width=6cm]{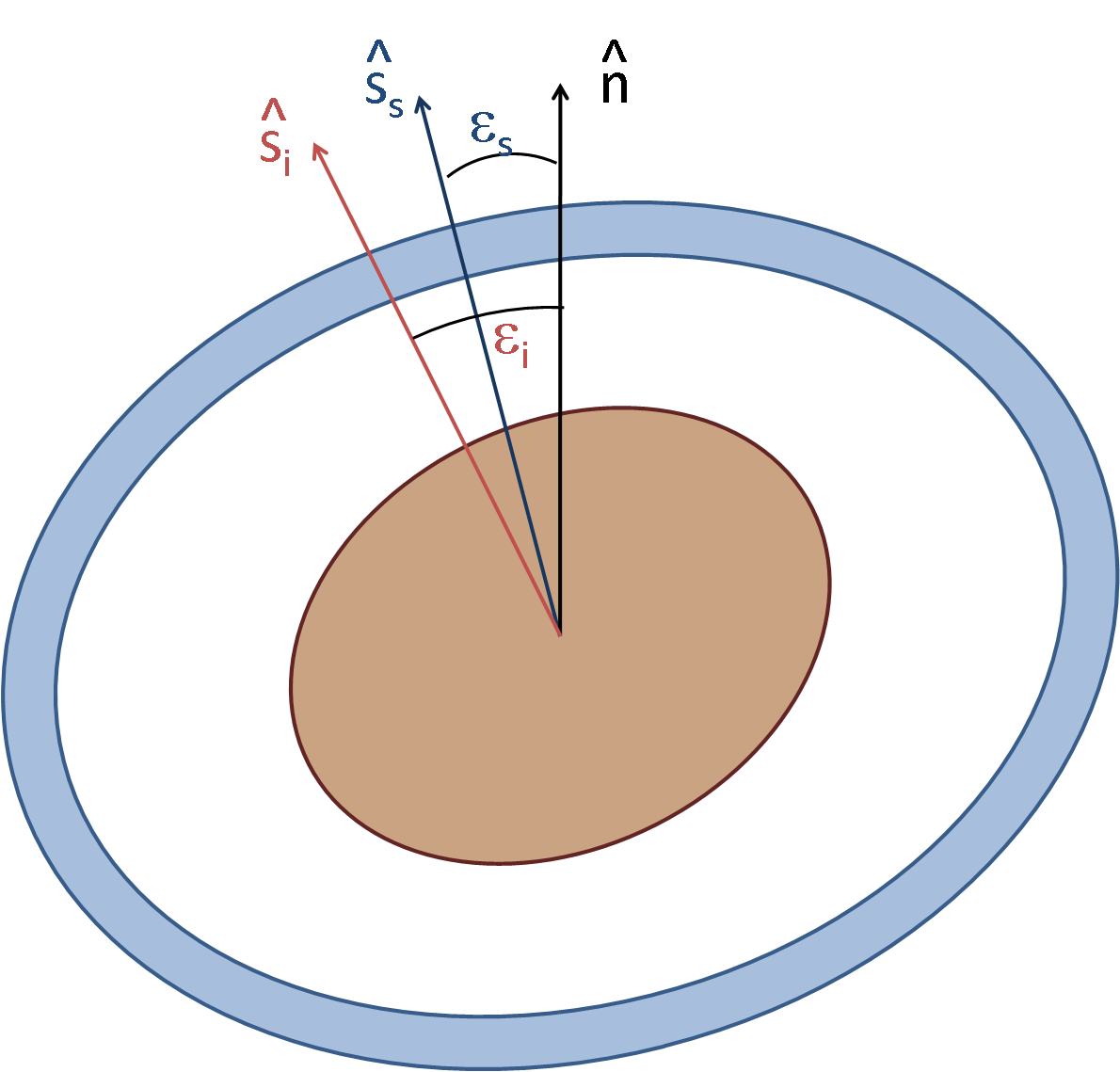}\\

\caption{Internal structure of a generic satellite, seen from the plane defined by the largest ($C$) and the smallest ($A$) moments of inertia. In the presence of an internal ocean, the rotation axes of the shell and of the interior can have different obliquities with respect to the normal to the orbital plane. }
\end{center}
\end{figure}

\begin{figure}[!htb]
\begin{center}
%Io\hspace{5cm} Europa\\
\includegraphics[width=14cm]{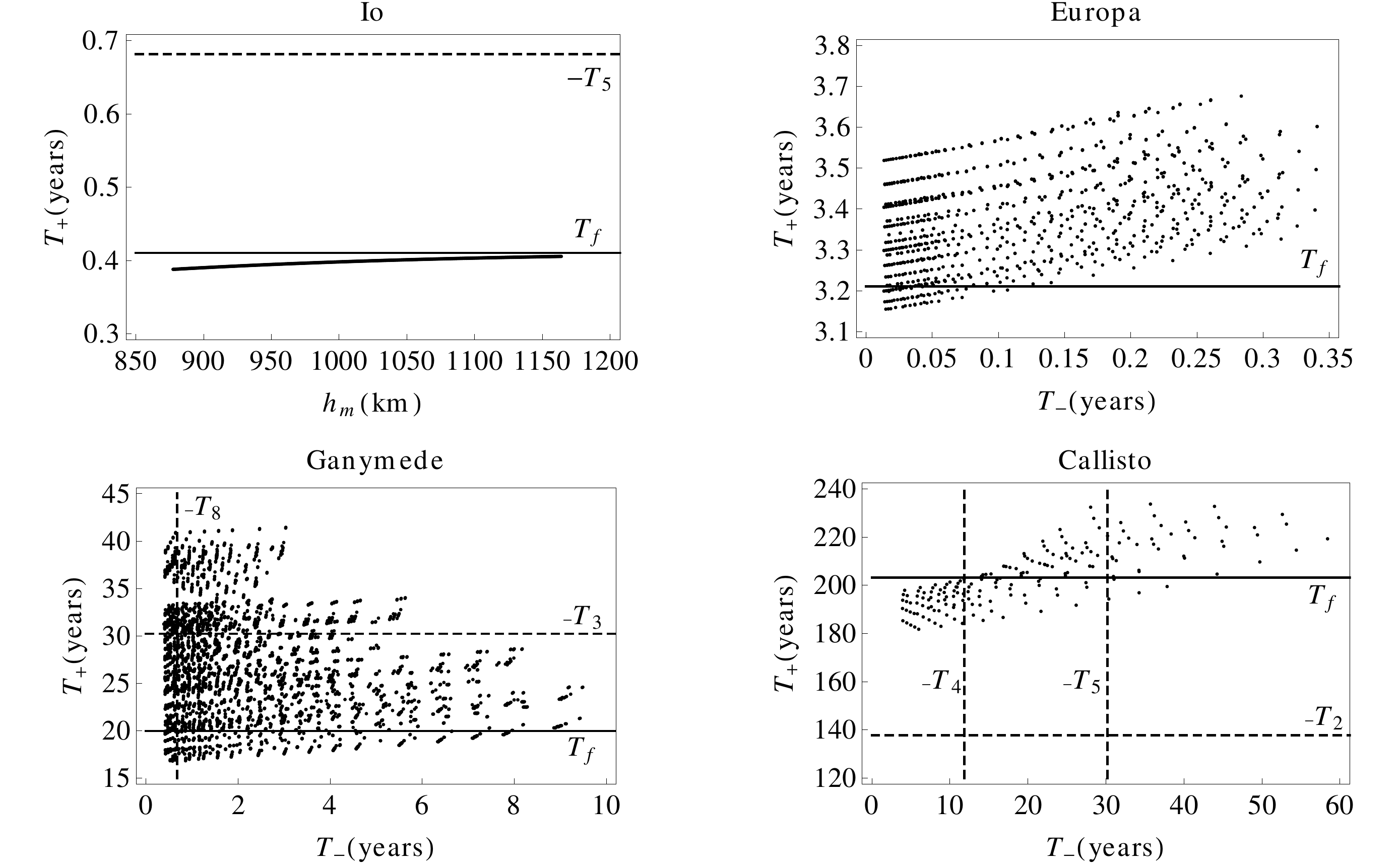}%\includegraphics[width=7cm]{tptm_Europa_quality.pdf}
%\includegraphics[width=8cm]{io_TpTm.png}
%\includegraphics[width=8cm]{europa_TpTm.png}\\
%Ganymede\hspace{5cm} Callisto\\
%\includegraphics[width=8cm]{gany_TpTm.png}\includegraphics[width=8cm]{calli_TpTm.png}\\

\caption{Periods of the free precession for the four Galilean satellites, in the case where they have an internal liquid layer. For Io, there is only one free period, plotted as a function of the mantle thickness $h_m$, for the different internal structure models considered. For the other satellites, the two free periods are plotted in function of each other, for the different internal structure models. For each satellite, the solid line corresponds to the period of the free precession of the solid case ($T_f$). Dashed lines correspond to the opposite periods ($-T_j$) of the orbital precession that may be close enough to one of the free periods to lead to a resonant amplification of the corresponding surface layer obliquity amplitude ($\varepsilon_{j,m}$ for Io and $\varepsilon_{j,sh}$ for the three other satellites) such as $\varepsilon_{j,m/sh}$ is of the same order of magnitude as $\varepsilon_{1,m/sh}$, as can be seen on Fig. 6.}
\end{center}
\end{figure}

\begin{figure}[!htb]
\begin{center}

\includegraphics[width=14cm]{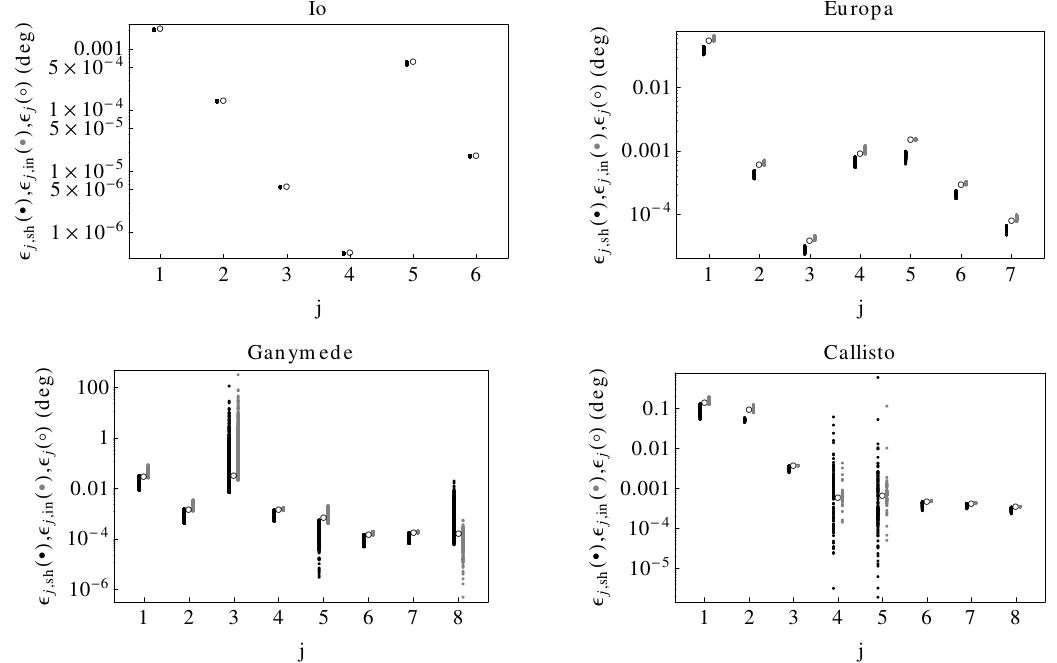}
%\includegraphics[width=8cm]{gany_epsshinj_old.png}
% Io\hspace{5cm} Europa\\
% \includegraphics[width=8cm]{io_epsshinj.png}\includegraphics[width=8cm]{europa_epsshinj.png}\\
% Ganymede\hspace{5cm} Callisto\\
% \includegraphics[width=8cm]{gany_epsshinj.png}\includegraphics[width=8cm]{calli_epsshinj.png}\\

\caption{Comparison of the obliquity amplitudes of the surface layer (black) and of the interior (gray) with those of the solid case (circles). }
\end{center}
\end{figure}

\begin{figure}[!htb]
\begin{center}
\includegraphics[width=14cm]{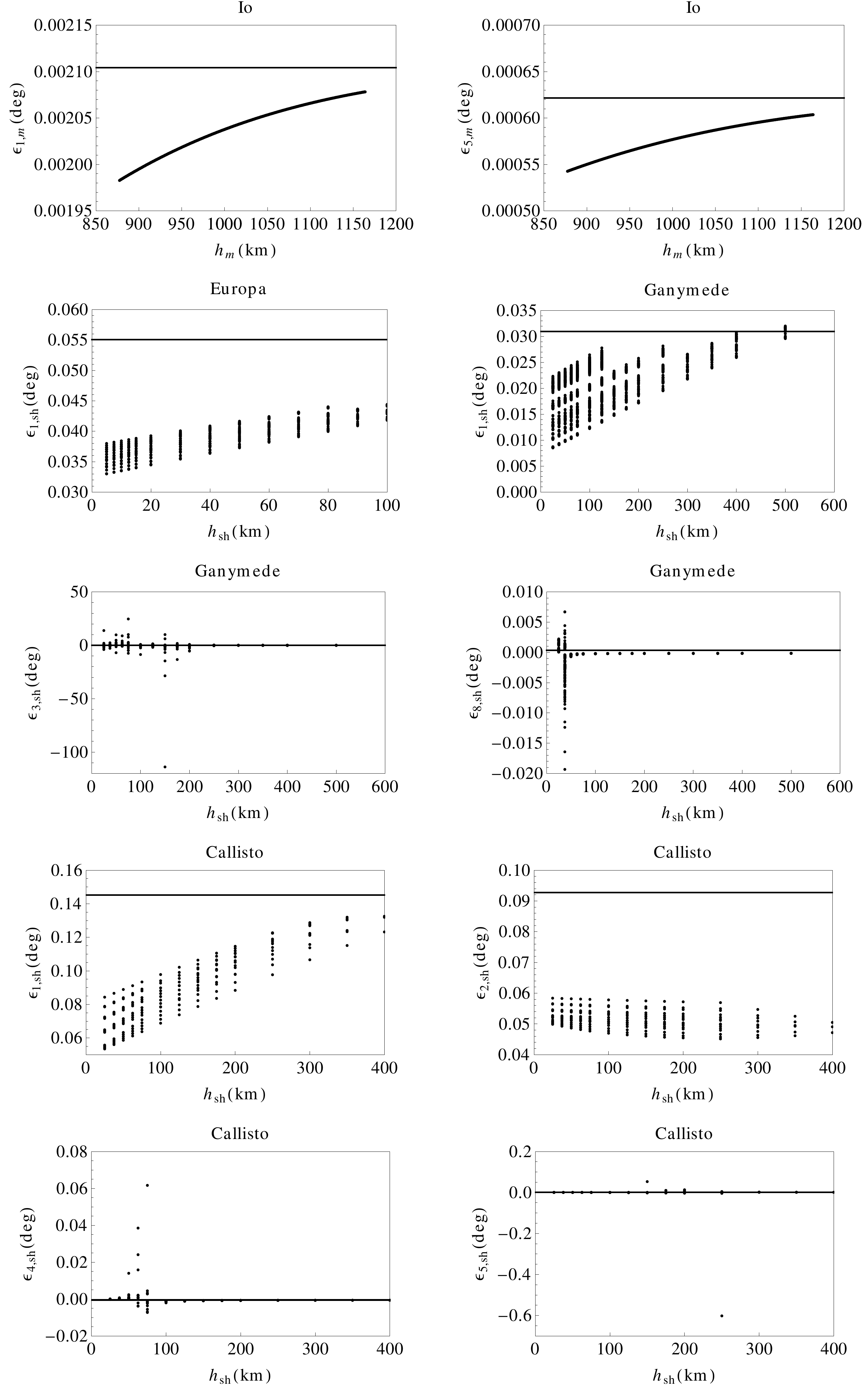}
% 
% Io\hspace{5cm} Europa\\
% \includegraphics[width=8cm]{io.png}\includegraphics[width=8cm]{europe.png}\\
% Ganymede\hspace{5cm} Callisto\\
% \includegraphics[width=8cm]{gany.png}\includegraphics[width=8cm]{calli.png}\\

\caption{Dependence of the dominant obliquity amplitudes with respect to the thickness of the surface layer, for the four satellites. Solid lines are for the entirely solid case.}
\end{center}
\end{figure}

\begin{figure}[!htb]
\begin{center}
\hspace{0cm}
\includegraphics[width=14cm]{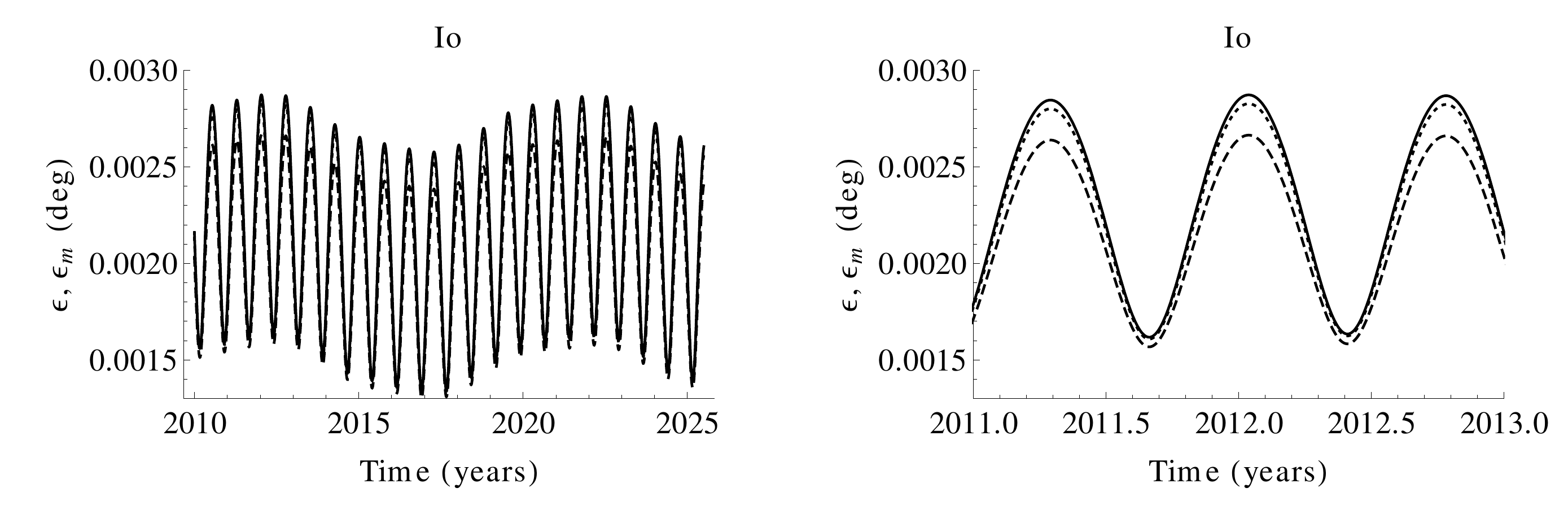}\\
\includegraphics[width=14cm]{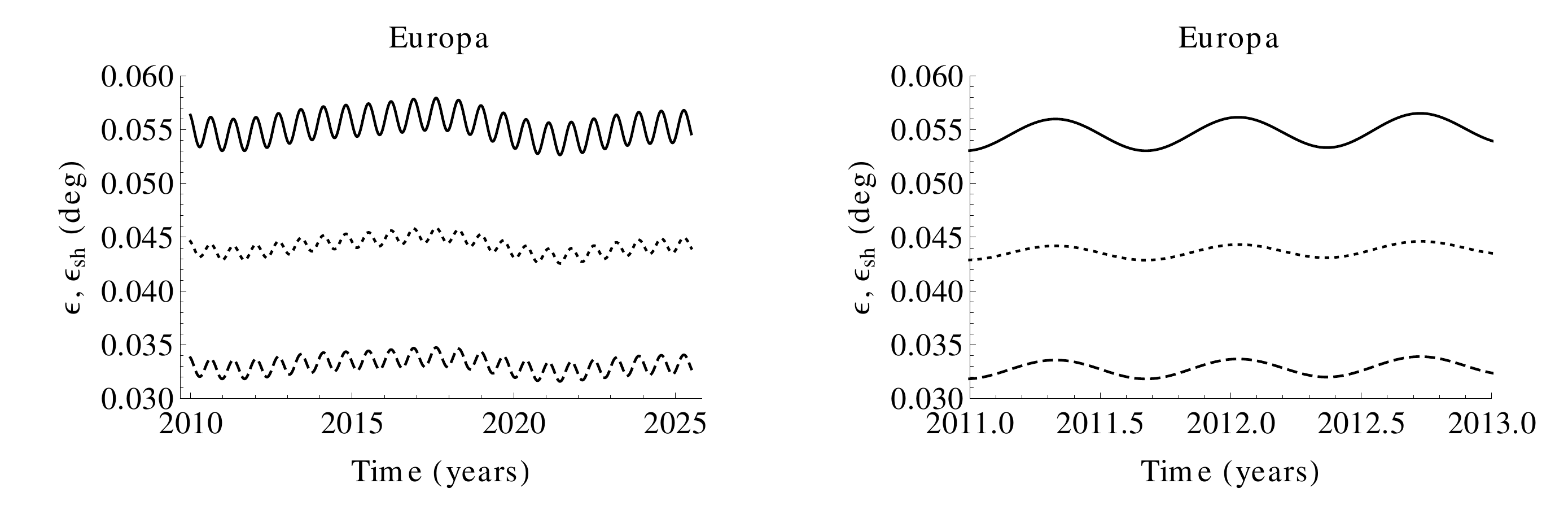}\\
\includegraphics[width=14cm]{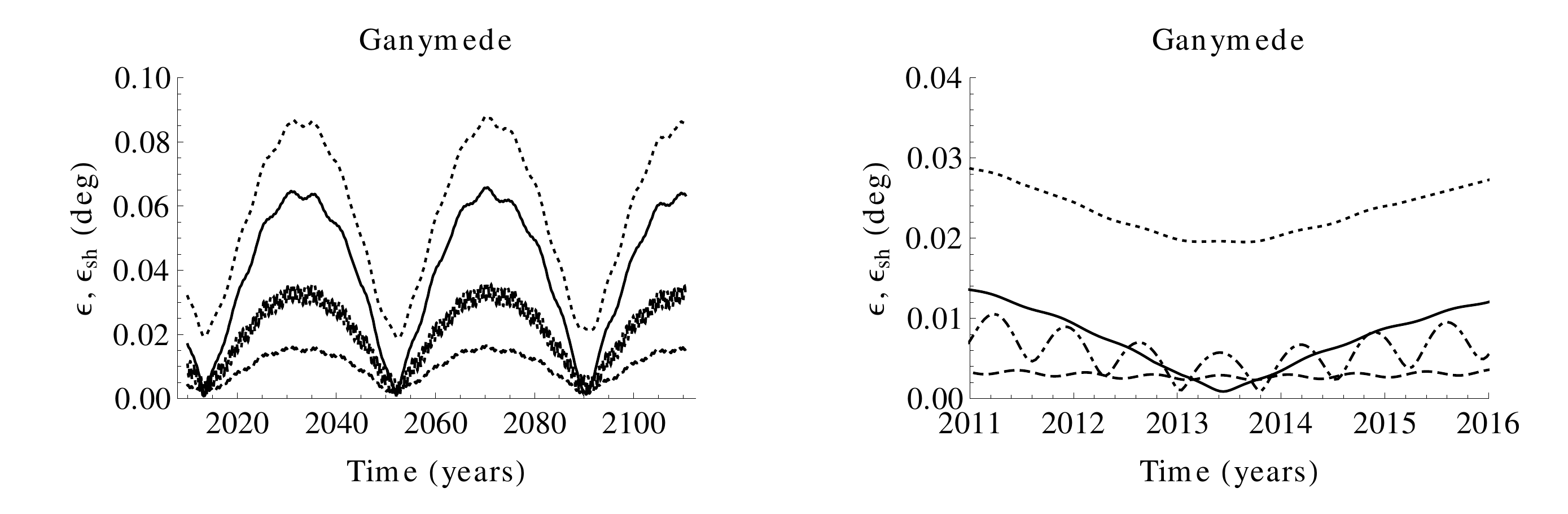}\\
\includegraphics[width=14cm]{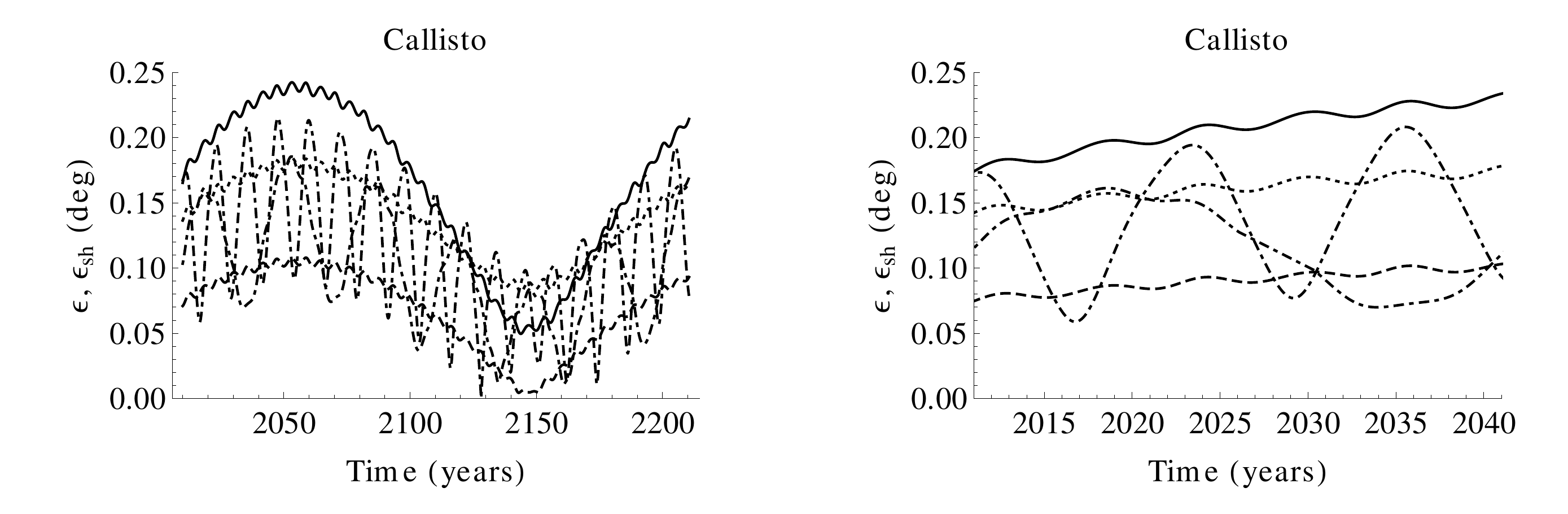}\\

\caption{Time evolution of the surface layer obliquity for the solid case (solid lines), for the internal structure models corresponding to the minimal and maximal values of $\varepsilon_{1,m/sh}$ (dashed lines and dotted lines, respectively), and for specific internal structure models which present a significant resonant amplification of another obliquity amplitudes (dot-dashed lines) than the previous ones. The time spans are quite long (left) or short (right), in order to be able to distinguish the different aspects of the time behavior.}
\end{center}
\end{figure}

\begin{figure}[!htb]
\begin{center}
\includegraphics[width=14cm]{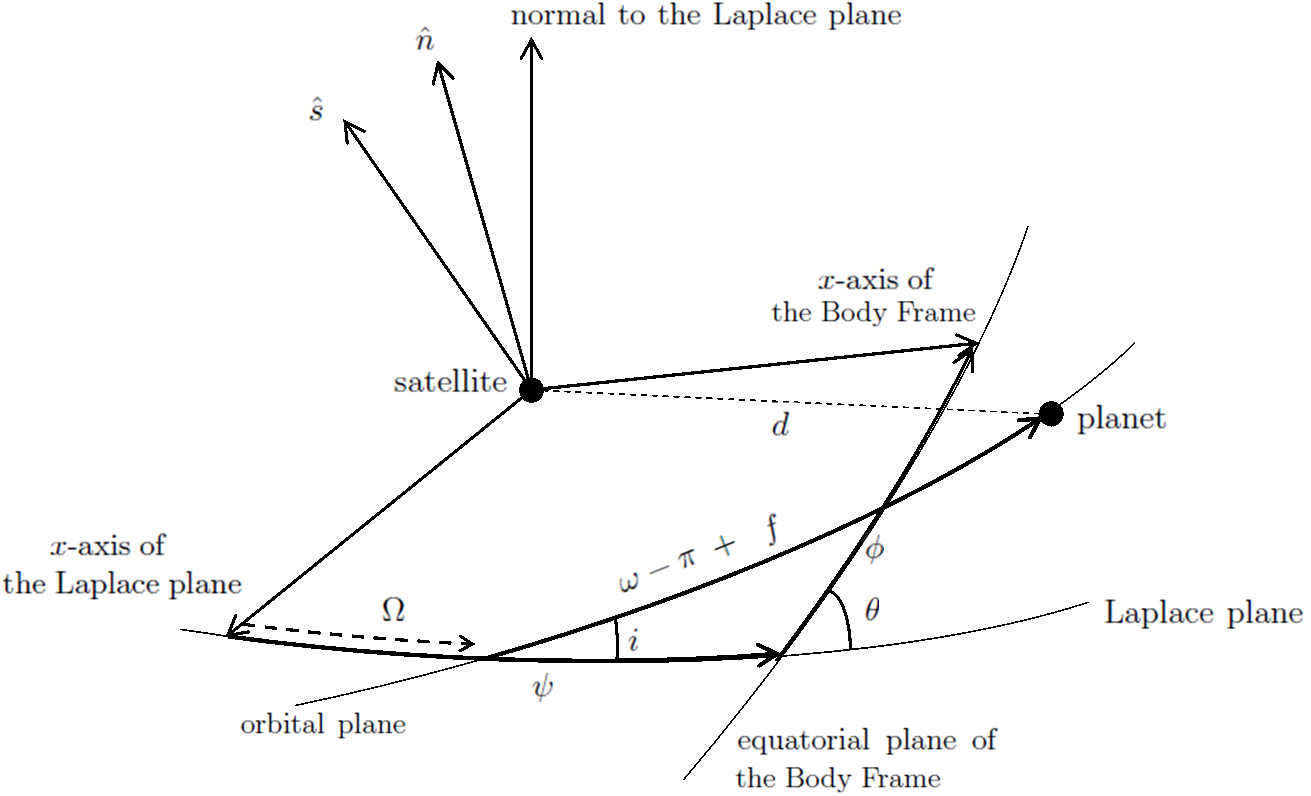}
% 
% Io\hspace{5cm} Europa\\
% \includegraphics[width=8cm]{io.png}\includegraphics[width=8cm]{europe.png}\\
% Ganymede\hspace{5cm} Callisto\\
% \includegraphics[width=8cm]{gany.png}\includegraphics[width=8cm]{calli.png}\\
Fig. 7
\caption{Orientation of the Body Frame of the satellite with respect to its orbital plane and to its Laplace plane.}
\end{center}
\end{figure}

\end{document}